\documentclass[10pt]{iopart}
\usepackage{iopams}

\usepackage{graphicx}
\usepackage{psfrag}
\usepackage{times}
\usepackage{amsthm}
\usepackage{amssymb}
\usepackage{color}

\usepackage{mathrsfs}

\DeclareSymbolFontAlphabet{\mathrsfs}{rsfs}
\DeclareMathAlphabet{\mathcal}{OMS}{cmsy}{m}{n}

\newcommand{\scri}{\mathrsfs{I}}

\newcommand{\be}{\begin{equation}}
\newcommand{\ee}{\end{equation}}

\def\pp{{\partial}}

\def\scri{{\mathscr{I}}}

\def\tt{{\tilde{t}}}
\def\tr{{\tilde{r}}}
\def\tx{{\tilde{x}}}
\def\tg{{\tilde{g}}}
\def\tl{{\tilde{l}}}
\def\hV{{\hat{V}}}

\def\dO{{\dot{\Omega}}}
\def\ddO{{\ddot\Omega}}

\def\om{{\,{^{\scriptscriptstyle \Omega}}}}
\def\omlam{{\,{^{\scriptscriptstyle \Omega\lambda}}}}

\def\nl{{ n\!\cdot\!l }}

\def\lKS{{l_{\textrm{\tiny KS}}}}
\def\lBL{{l_{\textrm{\tiny BL}}}}

\def\rbl{{r_{\textrm{\tiny BL}}}}
\def\rtrans{r_{\textrm{\tiny TR}}}
\def\lmax{l_{\textrm{\tiny MAX}}}

\def\Ric{ {\mathcal{R}} }

\begin{document}

\title[Hyperboloidal slices for the wave equation of Kerr-Schild metrics]{Hyperboloidal slices for the wave equation of Kerr-Schild metrics and numerical applications}

\author{Michael Jasiulek}

\begin{abstract}

We present new results from two open source codes, using finite differencing and pseudo-spectral methods for the wave equations in
(3+1) dimensions. We use a hyperboloidal transformation which allows direct access to null infinity and simplifies
the control over characteristic speeds on Kerr-Schild backgrounds. We show that this method is ideal for attaching
hyperboloidal slices or for adapting the numerical resolution in certain spacetime regions. As an example application,
we study late-time Kerr tails of sub-dominant modes and obtain new insight into the splitting of decay rates. The involved conformal
wave equation is freed of formally singular terms whose numerical evaluation might be problematically close to
future null infinity.

\end{abstract}

\section{Introduction}

In the recent past, since the work of \cite{Zenginoglu:2007jw,Zenginoglu:2008wc,Zenginoglu:2008uc}, hyperboloidal
slicings have been used frequently for numerical computations in general relativity as an elegant solution to the
outer boundary and the radiation extraction problem by including null infinity in the slicing
\cite{friedrich1983cauchy}.
The applications range from time-domain simulations of the scalar wave equation on Minkowski, Schwarzschild
\cite{Zenginoglu:2008uc,Zenginoglu:2010zm} and Kerr background \cite{Zenginoglu:2009hd,Racz:2011qu} to compute quasi-normal mode
oscillations, to study the late time behaviour of the solution, to study the influence of non-linear source terms, to compute the
scalar self-force of point particles orbiting a Schwarzschild black hole \cite{Vega:2011wf} and to compute gravitational
waveforms from extreme-mass-ratio inspirals \cite{Bernuzzi:2011aj,Zenginoglu:2011ik}. 
A new and interesting idea is to apply the framework to perturbation equations in the frequency domain \cite{Zenginoglu:2011jz}.
Recently, the use of hyperboloidal slices even allowed to evolve the Einstein equations in axisymmetry including null infinity,
see \cite{Rinne:2009qx} and references therein.

The spatial domain in the numerical solution of hyperbolic PDEs is typically truncated at a finite coordinate
distance where artificial outer-boundary conditions are imposed and the outgoing radiation of the test field
is extracted. This practice causes certain well-known conceptual and practical difficulties such as artificial
reflections on the outer-boundary which can destroy relevant features of the solution. In this context, the use of
hyperboloidal slicings offers an alternative by including the physical boundary $\scri^+$ on the numerical grid.

Many numerical applications require a specific coordinate system in a compact domain and it is necessary to
attach hyperboloidal coordinates at some transition point. In the original work of \cite{Zenginoglu:2007jw} this
was accomplished by introducing a transition zone, which required the fine-tuning of many parameters for the chosen transition
function and a disadvantageous hyperboloidal transformation which had a suitable asymptotic behaviour but caused the
outgoing characteristic speed to drop in the transition region. This led to numerical problems in subsequent works
\cite{Zenginoglu:2010zm,Zenginoglu:2009hd,Vega:2011wf}. In the recent work \cite{Zenginoglu:2011ik}, appearing
during the completion of this paper, the authors present a solution to the problem.

For this paper we developed a finite differencing (FD) and a pseudo-spectral (PS) code in (3+1) dimensions \cite{code:2013}
to test a new hyperboloidal transformation that simplifies the control over characteristic speeds for the 1st-order 
reduced wave equation
on Kerr-Schild backgrounds. Thereby we can avoid the above mentioned problem by requiring the outgoing characteristic
speed to be invariant under the compactifying hyperboloidal transformation. We demonstrate that by performing
a numerical comparison with the attached hyperboloidal slices as originally used in \cite{Zenginoglu:2007jw}.
As an example application we investigate the late-time decay rates in Kerr at the horizon, finite radii and null
infinity and study the $m$-dependence of the splitting of certain sub-dominant modes which was found in the numerical studies
of \cite{Zenginoglu:2009hd,Racz:2011qu}.
More technically, we remove formally singular terms that appear in the conformal wave equation that might 
numerically be problematic to evaluate at the outer boundary $\scri^+$, where the conformal factor vanishes.
Our framework is general enough to cover metrics of the Kerr-Schild form. Thus, our approach could be applied in
the numerical analysis of quasi-normal modes in Vaidya \cite{Chirenti:2011rc}, and more generally, in time-dependent
non-axisymmetric Kerr-Schild spacetimes.

The infra-structure and implementations we present here are comprehensive, allowing standard coordinates in the
spacetime interior in (3+1), smooth matching of hyperboloidal slices and covering the large class of Kerr-Schild
metrics with FD and PS techniques. A comparable work is \cite{Zenginoglu:2010zm} in which Minkowski and Schwarzschild
backgrounds are treated on hyperboloidal slices in (3+1) with PS techniques using the aforementioned non-optimal
transition zone.

The paper is organised in the following way. In Sec. (\ref{sec:wave-Kerr}) we introduce our notation and convention of
Kerr-Schild metrics and the scalar wave equation. At next Sec. (\ref{sec:hyperboloidal-method}) we briefly explain
the hyperboloidal method as well as the particular hyperboloidal transformation we employed. Then we compare the
resulting characteristic coordinate speeds and the scalar curvature for matched hyperboloidal slices with other
methods. In section (\ref{sec:wave-hyperboloidal}) we apply our method to solve the scalar wave equation with
finite differencing, explain implementational details and compare different matching methods numerically
in Sec. (\ref{sec:num}). We then proceed to explain the pseudo-spectral code and use this on a single hyperboloidal domain
(no matching) to study polynomial decay rates of sub-dominant modes on Kerr.

Note that we use a tilde as in $\tr$ or $\tg^{\mu\nu}$ to denote compatified hyperboloidal coordinates and components
of tensors. Conformally rescaled tensors are denoted by a superscript $\Omega$ as in $\om l^\mu=l^\mu/\Omega$
or $\om g^{\mu\nu}=g^{\mu\nu}/\Omega^2$. Derivatives wrt the Kerr-Schild radius $r$ are denoted by a prime like
$dh/dr=h'$ and wrt the rescaled radius $\tr$ by a dot $dh/d\tr =\dot{h}$.

\section{Wave equation in Kerr-Schild coordinates} \label{sec:wave-Kerr}

\subsection{ Kerr-Schild metrics and Kerr metric }

Kerr-Schild metrics are of the general form 
\begin{equation} \label{eq:ks-metric}
   g^{\mu\nu} = \eta^{\mu\nu} - 2 V(t, x^i )\,  l^\mu l^\nu , 
\end{equation}
where $\eta^{\mu\nu}= \textrm{diag}( -1,1,1,1 )$ is the flat metric, $( t,x^i )$ are Kerr-Schild coordinates and $l^\mu$ is a null vector
wrt $\eta^{\mu\nu}$ ($\Rightarrow$ $g^{\mu\nu}l_\mu l_\nu=0$) of the form
\begin{equation} \label{eq:lknull}
   l^\mu = (-1,\, l^i ),  \quad
   k^\mu = ( 1, l^i ) - 2 V\, l^\mu
\end{equation}
and $k^\mu$ is the second null vector normalized to $k^\mu l_\mu = 2$ ( $\Leftrightarrow l^i l^j \delta_{ij}=1$ ). $l^i$ and $V$ \footnotemark are free functions that
characterise the Kerr-Schild metric in question.
\footnotetext{ \label{fn:asymptotic_flatness}
We assume asymptotic flatness such that $V = M/r( 1+ V_1/r + \mathcal{O}(1/r^2))$ and $l^i n_i = 1 + \nl_1/r + \mathcal{O}(1/r^2)$, $n^i=x^i/r$. 
In Kerr for example $V = M/r ( 1+ \mathcal{O}(1/r^2))$ and $\nl=1 - a^2/2 (1-(n_z)^2)/r^2 - \mathcal{O}(1/r^4)$.
}

In Kerr with mass $M$ and angular momentum $a\, M$ we have 
\begin{equation}
   V(x^i) = M \frac{ \rbl }{ s  } \quad ( = \frac{M}{2} \nabla_\mu l^\mu ),
\end{equation}
where $s := \sqrt{ ( r^2 - a^2 )^2 + (2 a z)^2 }$, $\rbl = \frac{1}{\sqrt{2} }  \sqrt{ r^2 - a^2 + s }$ is the
Boyer-Lindquist radius and $r=\sqrt{x^2+y^2+z^2}$ the Kerr-Schild radius, and 
\begin{equation}
  l^i = \left( \frac{ r_{\textrm{\tiny BL}} x + a\, y }{ \rbl^2 + a^2 }, \frac{\rbl y - a\, x}{ \rbl^2 + a^2 }, \frac{z}{\rbl} \right).
\end{equation}

Then $(-l^\mu)$ is the future directed ingoing principal null vector and $k^\mu$ the outgoing one.  
With our sign convention 
\begin{equation}
 \nl = + \frac{\rbl}{r},
\end{equation}
where $\nl:=n_\mu l^\mu$, $n_i=n^i=x^i/r$ is the radial unit normal.
Note that $l^\mu$ is affinely parametrised, i.e. $l^\mu \nabla_\mu l^\nu=0$. Sometimes it can be useful to work with $\hat{l}^\mu := \sqrt{ 2 V} l^\mu$. 
Then $\hat{l}^\mu \, \nabla_\mu \hat{l}^\nu = \phi\, \hat{l}^\nu = \pp_l V \,l^\nu  = M \frac{a^2-r^2}{s^2} \sqrt{\frac{1 }{ 2V }} \, \hat{l}^\nu $.  

\subsection{ Wave equation in 1st-order form }

The hyperboloidal method, explained in the next section, includes a conformal rescaling of the metric. Therefore, we consider the conformal
wave equation instead of $\Box \Psi=0$ \footnotemark. 
\footnotetext{
The scalar wave equation $\Box \psi =0$ is per se not conformally invariant, i.e. a rescaled solution $\om \psi := \psi/\Omega$ 
is in general not a solution of the wave eq. of the rescaled metric $\om g^{\mu\nu}:= g^{\mu\nu}/\Omega^2 $, see for example appendix D of \cite{Wald:1984rg}.  
Note that $\Ric=0$ holds for Kerr-Schild metrics where $l^\mu$ is geodesic like in Kerr or Vaidya spacetimes.
}

We reduce the conformal wave equation
\begin{equation} \label{eq:wave}
   (\Box - \frac{1}{6} \Ric ) \psi = g^{\mu\nu} \pp_\mu \psi_\nu - \Gamma^\mu(g) \psi_\mu - \frac{1}{6} \Ric \psi = 0
\end{equation}
to 1st-order form by introducing the new variables $\psi_\nu := \pp_\nu \psi$, where $\Gamma^\lambda(g) := g^{\mu\nu} \Gamma^\lambda_{ \mu\nu }$
are the contracted Christoffel symbols and $\Ric$ the scalar curvature of $g^{\mu\nu}$.
In general $\Gamma^\mu = ( \nabla_\mu \hat{l}^\mu + \phi )\, \hat{l}^\mu$ holds for metrics of the form (\ref{eq:ks-metric}).
In Kerr it can be shown that 
\begin{equation} \label{eq:gammaKS}
   \Gamma^\mu(g) = \frac{ 2 M}{ s} l^\mu.
\end{equation}
This leads to the following system of evolution equations for the variables $\{ \psi_{\nu=0,1,2,3}, \psi \}$: 
$\pp_0 \psi = \psi_0$, $ \pp_0 \psi_j = \pp_j \psi_0$ and 
\begin{eqnarray} \label{eq:waveKerr}
   -g^{00} \pp_0 \psi_0 &=&  2 g^{0i} \pp_i \psi_0 + g^{ij} \pp_i \psi_j - \Gamma^\mu(g) \psi_\mu - \frac{1}{6} \Ric \psi \label{eq:KSevolvepsi0} 
\end{eqnarray} 
which is symmetric hyperbolic. For finite-differencing 
implementations of (\ref{eq:waveKerr}) the following flux-conservative form is more convenient for long-term stability of the
numerical evolution
\begin{eqnarray} \label{eq:waveKerr2}
   -g^{00} \pp_0 \psi_0 &=& g^{0i} \pp_i \psi_0 + \frac{1}{\sqrt{g}} \pp_i F^i - \frac{1}{6} \Ric \psi \quad \textrm{where} \\
   F^i &:=& \sqrt{g} \, ( g^{0i} \psi_0 + g^{ij} \psi_j ). \label{eq:Fi}
\end{eqnarray}
where finite differences are taken of $\psi_0$ and $F^i$ instead of differentiating the evolution variables $\psi_0$ and $\psi_i$ 
directly. This, in turn, is more efficient if spectral methods are used to compute spatial derivatives and eq. (\ref{eq:waveKerr})
is preferred in that case.

\section{Hyperboloidal method} \label{sec:hyperboloidal-method} 

Having explained the wave equation on Kerr-Schild backgrounds we are now ready to describe the hyperboloidal method
in the following section based on \cite{Zenginoglu:2007jw}. We explain the steps necessary to transform a given
spacetime metric $g^{\mu\nu}$ in coordinates $(t,x^i)$ and its $t$-slicing ($t=\textrm{const}$ hypersurfaces)
to a conformal metric $\om \tg^{\mu\nu} $ in radially compactified hyperboloidal coordinates $(\tt,\tx^i)$ and
its hyperboloidal $\tt$-slicing.
It is often desirable to leave the coordinates $(t,x^i)$ untouched in the spacetime interior, \emph{inner domain}, and
attach the \emph{hyperboloidal domain} smoothly at some transition point $r_{\textrm{\tiny TR}}$.
The hyperboloidal method we employ consist of three transformations, a radial compactification and a conformal rescaling of the metric 
\begin{eqnarray}
   \tx^j &=& \Omega(r)\, x^j                \quad  \textrm{with}  \quad \Omega =0,\,\, \dO \neq 0 \,\, \textrm{at} \,\, \scri^+     \label{eq:compact}  \\
   ^{\Phi} g_{\mu\nu} &=& \Phi^2 g_{\mu\nu} \quad\,\,  \textrm{with}  \quad \Phi \sim \Omega \,\, \textrm{at} \,\, \scri^+ \label{eq:gomega}
\end{eqnarray}
where $\dO:=d\Omega/d \tr$ and we assume $\Omega$ and $\Phi$ non-negative. For simplicity we set $\Omega \equiv \Phi$
in the following, see \ref{appendixB} for $\Omega\neq \Phi$. The conformal transformation extends the spacetime to include $\scri^+$ as its boundary \cite{friedrich1983cauchy}
such that the conformal metric is regular there.
For constant time slices to intersect $\scri^+$ we require the additional transformation 
\begin{equation}  \label{eq:ttilde}
  \tt = t - h(r) \quad \textrm{with} \quad  -\frac{I - 2 \nl V  }{ 1- 2 \nl^2 V } < h' < \frac{I + 2 \nl V  }{ 1- 2 \nl^2 V }
\end{equation}
where the height function $h(r)$ possesses a carefully chosen asymptotic singular behaviour to allow $\tt$-slices to penetrate $\scri^+$,  
similar to time transformations from Schwarzschild slicing to horizon penetrating slicings, see for figure (2) in \cite{Zenginoglu:2009ey} for illustration purposes.
The restrictions on $h(r)$ in (\ref{eq:ttilde}) are necessary for the hyperboloidal slices to be spacelike, where $h':=dh/dr$ and
$I:=\sqrt{ 1+2V(1-\nl^2) }$. 

\subsection{Controlling the coordinate speed of in/outgoing characteristics }

In numerical applications of the hyperboloidal method, like for the hyperbolic system (\ref{eq:waveKerr}), it is often
necessary to limit the ratio $\frac{\Delta t}{\Delta r}$ to obtain a stable numerical evolution, where $\Delta r$ is
the spatial grid spacing and $\Delta t$ the time step. This is the case, if the method of lines (MoL) and explicit
time integration is used. Then the CFL-condition must hold $c(r)\, \frac{\Delta t}{\Delta r} \le \nu_{\textrm{\tiny
CFL}}$, where $c(r)$ is the coordinate speed of characteristics of the solution, explained in the following,
and $\nu_{\textrm{\tiny CFL}}$ the so called Courant number that is independent of the solution. Therefore,
an efficient numerical computation is limited by $\max\, c(r)$ and computing time may be wasted in regions where
$c(r)< \max\, c(r)$.  On the other hand it may be useful to obtain more temporal resolution in certain regions,
for example to resolve a particular feature of the solution or to increase the accuracy there.
In the following we bring the height function derivative $h'(r)$ in a particular form which allows to directly
specify the outgoing characteristic speed $c(r)$ as a function of radius in the hyperboloidal transformation,
i.e. $h'(r) \rightarrow h'(r,c(r))$ independently of $\Omega$. We also show how to use this feature for attaching
hyperboloidal slices to an inner domain \footnotemark.
\footnotetext{
As mentioned in the introduction in the recent work \cite{Zenginoglu:2011ik} this problem is solved in a slightly different way.
The authors do not explicitly consider the characteristic speeds but guarantee a smooth transition of inner and hyperboloidal 
coordinates through $t_{\textrm{\tiny BL}} - r^*_{\textrm{\tiny BL}} = \tt - \tr $, see eq.(7) of \cite{Zenginoglu:2011ik}, where 
$t_{\textrm{\tiny BL}}$, $r^*_{\textrm{\tiny BL}}$ are the Boyer-Lindquist time and tortoise radius.}
%

The in/outgoing characteristic speeds in the direction $n_i$ of a hyperbolic system of the form $\pp_t U = A^i \pp_i U + B U$, where we
assume $U$ is vector-valued and $A^i, B$ are matrices, are given by two eigenvalues of $A^i n_i$. For the systems (\ref{eq:waveKerr}), (\ref{eq:waveKerr2}) they are
\begin{eqnarray}
 c^{\pm}(g):&=& \left( -g^{0n} \pm \sqrt{ (g^{0n})^2 -g^{00}\, g^{nn} } \right) / ( - g^{00} ), \label{eq:cpm_g} \\ 
            &=& \frac{\pm I - 2 V \nl }{ 1+2V} \quad \buildrel { \nl \to 1} \over \longrightarrow  \quad  \frac{+1-2V}{1+2V}; -1 \nonumber 
\end{eqnarray}
where $g^{0n}=g^{0i}n_i$. We considered the limit $\nl\rightarrow 1$ because in most parts of asymptotically flat spacetimes,
see footnote \ref{fn:asymptotic_flatness}, the two unit normals are almost aligned $\nl \approx 1$, e.g. in extremal Kerr for $r>1.2$.
Instead of proceeding with the straightforward but long algebraic operations to simplify $c^{\pm}(\om \tg)$, we proceed in a more intuitive way by considering
the characteristic speeds of the 1st-order equation 
\begin{equation} \label{eq:dl_psi}
   l^\mu \pp_\mu \psi = 0
\end{equation}
which applies to the characteristics of the PDEs (\ref{eq:waveKerr}), (\ref{eq:waveKerr2}). Characteristics propagate on 
null rays of the background along which the solution is constant. The radial characteristic speeds of eq. (\ref{eq:dl_psi}) along the null vectors 
$(-l^\mu),\, k^\mu$ eq. (\ref{eq:lknull}) are    
\begin{eqnarray} \label{eq:ckl}
 c^{(k)}  &= k^i n_i / k^0 = \nl \frac{ 1-2V }{ 1+2V } & \quad \buildrel { \nl \to 1 } \over \longrightarrow \quad c^{(k)} \equiv c^+(g), \\  
 c^{(l)}  &= l^i n_i / l^0 = - \nl                     & \quad \buildrel { \nl \to 1 } \over \longrightarrow \quad c^{(l)} \equiv c^-(g).
\end{eqnarray}

\begin{figure}
   \begin{minipage}[b]{0.5\linewidth}
   \begin{center}
      \includegraphics[width=1.0\columnwidth,clip]{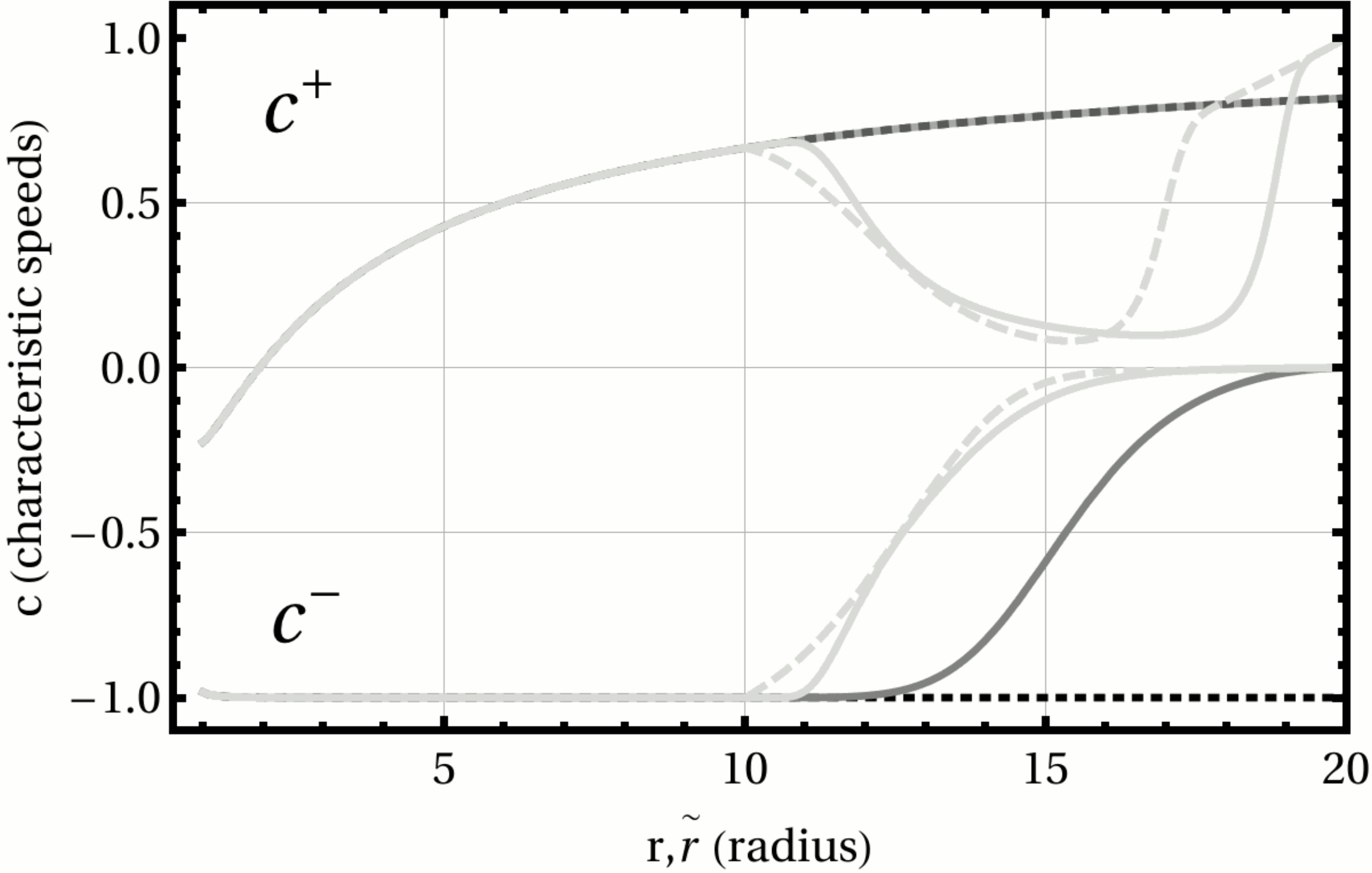}
   \end{center}
   \end{minipage}
   \begin{minipage}[b]{0.5\linewidth}
   \begin{center}
     \includegraphics[width=1.0\columnwidth,clip]{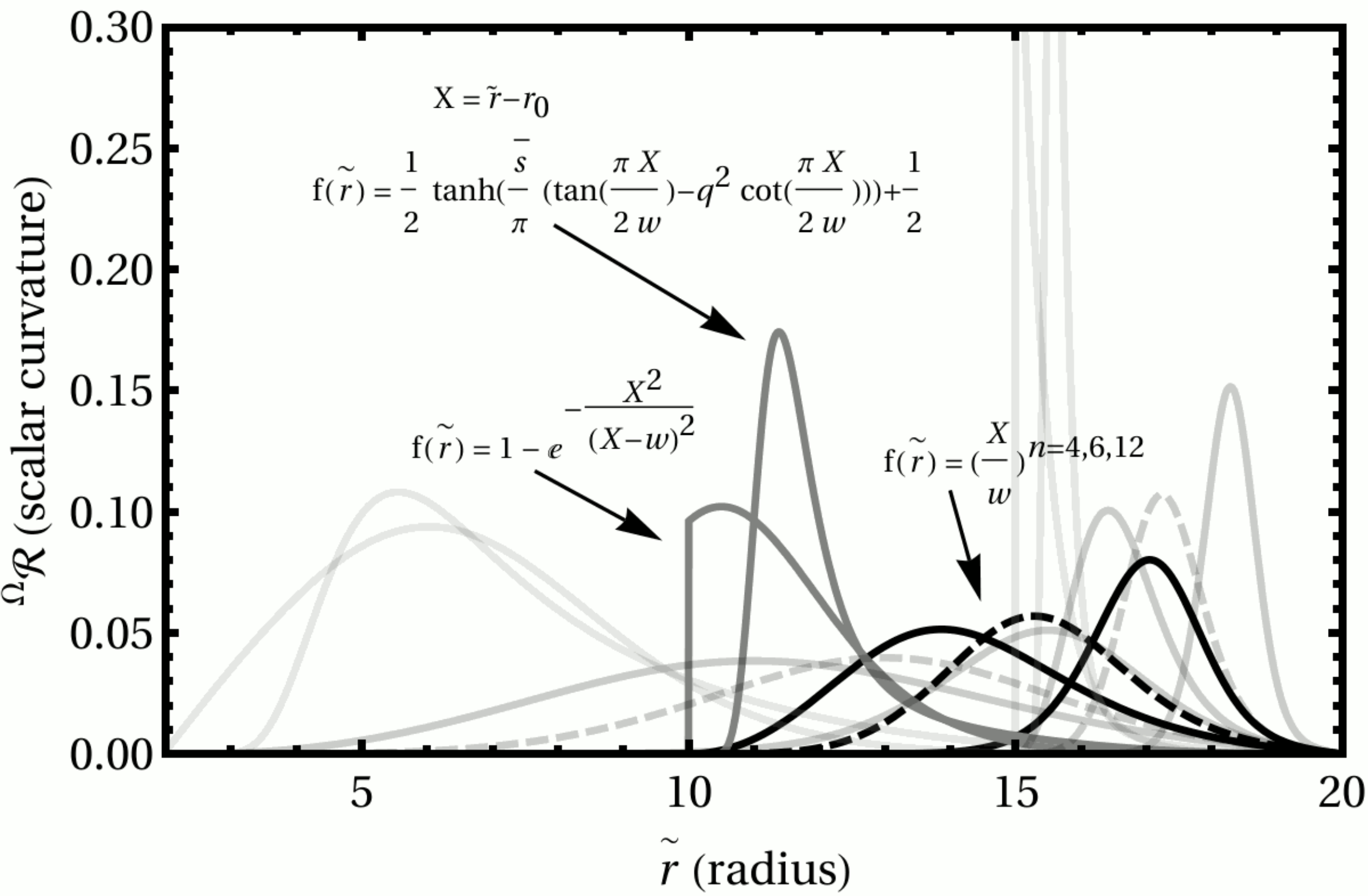}
   \end{center}
   \end{minipage}
\caption{
    Left: Characteristic speeds $c^\pm$ of the systems (\ref{eq:waveKerr}), (\ref{eq:waveKerr2}) for Kerr
          $a=0.5$ on $t$-slices (black dotted) $r\in[2;20]$ and compactified hyperboloidal $\tt$-slices with
          $\tr|_{\scri^+}=20$ attached at $\rtrans=10$ for the hyperboloidal matching condition (\ref{eq:matching_condition2}) (gray) and for (\ref{eq:hext}) with $C=20$
          of Refs. \cite{Zenginoglu:2007jw,Zenginoglu:2010zm,Zenginoglu:2009hd,Vega:2011wf} (light gray) for transition
          functions $f$-tanh-tan (light g. solid) with $q=1$,$\bar{s}=2$ \cite{Zenginoglu:2010zm,Vega:2011wf} and $f$-exp (light g. dashed)
          \cite{Zenginoglu:2009hd,Zenginoglu:2007jw}. \\ 
   Right: $\om\Ric(\tr)$ for the conformal factors (transition functions $f(\tr)$) of 
          Refs. \cite{Zenginoglu:2007jw,Zenginoglu:2010zm,Zenginoglu:2009hd,Vega:2011wf} (gray)
          and our choice (black, $k=6$ dashed).
          The same five curves are shown in two light gray groups for shifted transition points $\rtrans=2;15$.}
\label{fig:coord_speeds}
\label{fig:scal-curv-trans}
\end{figure}

Under the compatifying hyperboloidal transformations (\ref{eq:compact}),(\ref{eq:ttilde}) $c^{(k)}, c^{(l)}$ change as
\begin{eqnarray} 
   c^{(\tilde{k})} &= \nl \frac{ 1-2V }{ 1+2V }\, \frac{ \Omega^2\, L} { 1 - h'\, \frac{1-2V}{1+2V} \nl } 
    & \quad \buildrel { \nl \to 1 } \over \longrightarrow \quad c^{(\tilde{k})} \equiv c^+(\tg),  \label{eq:ctilde_k}  \\
   c^{(\tilde{l})} &=  -\nl \, \frac{\Omega^2\, L}  {1+h' \nl} 
    & \quad \buildrel { \nl \to 1 } \over \longrightarrow \quad c^{(\tilde{l})} \equiv c^-(\tg), \label{eq:ctilde_l} 
\end{eqnarray}
where $L:= \frac{\tr'}{\Omega^2} = \frac{1 }{\Omega - \dO \tr } $.
From eq. (\ref{eq:ctilde_l}) we can see that the conditions (\ref{eq:compact}) on $\Omega,\dO$ make $c^{(\tilde{l})}$ vanish at $\scri^+$ (no ingoing characteristics).
Eq. (\ref{eq:ctilde_k}) shows that to make $c^{(\tilde{k})}$ non-vanishing at $\scri^+$, $h'$ has to be of the form
\begin{eqnarray} \label{eq:matching_condition}
   h' &=& \frac{1+2V}{1-2V} \left( 1- \Omega^2 L \, H \right) \,\, \buildrel { h'=h'(r) } \over \longrightarrow \,\,
          \frac{1+2\hV(r)}{1-2\hV(r)} \left( 1- \Omega^2 L \, H(r) \right),
\end{eqnarray}
where $\hV(r)$ is $V(x^i)$ along some arbitrary direction or some approximation. For practical purposes it is enough to consider
the leading order term of $V(x^i)$, i.e. $\hV(r)= 2M/r$, as we do in the following.
$H$ is a free function in the range from zero to one to modify the
coordinate speeds in the hyperboloidal domain. The spacelike condition eq. (\ref{eq:ttilde}) is guaranteed by $h'$ of the above form, since $L H>0$
by definition \footnotemark.
\footnotetext{
One may object that for $h'$ of (\ref{eq:matching_condition}) in the limit $\Omega \rightarrow 0$ $\frac{I+2\nl V }{1-2\nl^2 V} < h'$. By expanding this
inequality in $\Omega$ at $0$ it becomes clear that this can only be if $L H<0$.
}
A simple choice might be $H=1$. However, $H$ allows direct access to the characteristic speed $c^{(\tilde{k})} = (1-2V)/(1+2V)\, 1/H$ eq. (\ref{eq:ctilde_k}) and it is more convenient to require
\begin{equation} \label{eq:matching_condition2} 
   c^{(\tilde{k})} = c^{(k)} \quad \buildrel {\nl \to 1 } \over \Longrightarrow \quad H(\tr) = \frac{1-2\hV(r)}{1+2\hV(r)} \cdot \frac{1+2 \hV(\tr) }{1-2\hV(\tr) }.
\end{equation}
This condition is the core of our hyperboloidal method. It ensures that the outgoing characteristic speed
is unchanged under the compatifying hyperboloidal transformation, see Fig. (\ref{fig:coord_speeds}) (left) (gray
curve overlapping black dotted curve), independent of the choice for $\Omega$. We set $\Omega$ in the following to be
\begin{equation} \label{eq:omega-choice}
 \Omega(\tr) = 1-f(\tr) \quad \textrm{with} \quad f(\tr)= \left( \frac{X}{w} \right)^k\, \Theta(X),\, X=\tr-\tr_{\textrm{\tiny TR}},
\end{equation}
where $\tr_{\textrm{\tiny TR}}$ is the transition point and $w$ the width of the hyperboloidal domain.

In the original work \cite{Zenginoglu:2007jw} and also adapted by \cite{Zenginoglu:2010zm,Zenginoglu:2009hd,Vega:2011wf} the height function derivative 
was set to
\begin{equation} \label{eq:hext}
   h'= f\, H_{\textrm{\tiny ext}} = f\, \left( 1+ \frac{4M}{r} + \frac{(8M^2-C^2)}{r^2} \right)
\end{equation}
where $C$ is a parameter and $f$ ranging from zero to one a transition function and the conformal factor was set to $\Omega(\tr)=1-f(\tr)
\frac{\tr}{S},\, S=\tr|_{\scri^+} $. $H_{\textrm{\tiny EXT}}$ has the correct asymptotic singular behaviour, where $c^+|_{\scri^+}=S^2/C^2$
but the characteristic speed in the hyperboloidal domain is left uncontrolled in the transition zone and depends on the details of $f$ 
(as the spacelike condition (\ref{eq:ttilde})).
In this case $c^{(\tilde{k})}$, see eq. (\ref{eq:ctilde_k}), drops like $\Omega^2$ before $h'$ in the denominator sets in to result
in a non zero $c^{(\tilde{k})}$ asymptotically. This effect is apparent in Fig. (\ref{fig:coord_speeds}) (left) (light gray, light gray
dashed), where we plotted $c^\pm$ (\ref{eq:cpm_g}) for our choice (\ref{eq:matching_condition}), (\ref{eq:matching_condition2})
and for (\ref{eq:hext}) as in Refs. \cite{Zenginoglu:2007jw,Zenginoglu:2010zm,Zenginoglu:2009hd,Vega:2011wf}.

\section{Wave equation on hyperboloidal slicing} \label{sec:wave-hyperboloidal}

In the following we derive the explicit expressions for $\om \tg^{\mu\nu}$, $\om \Ric$ and $\Gamma^\mu(\om\tg)$
that appear in the conformal wave eq.(\ref{eq:wave}) under the compatifying conformal hyperboloidal transformations
(\ref{eq:compact}),(\ref{eq:gomega}),(\ref{eq:ttilde}). 
We identify all formally singular terms which could constitute potential obstacles for numerical
evaluations at the outer boundary $\scri^+$. 

\subsection{Transformation of the Kerr-Schild metric $g^{\mu\nu}$ and $\Ric$, $\Gamma^\mu$}

The metric components $g^{0\mu}$, $g^{ij}$ and the determinant of $g^{\mu\nu}$ transform 
in a straightforward manner, divergent terms cancel trivially, the expressions are given in \ref{appendix}.
The regularity of 
\begin{equation} \label{eq:gt00}
   \om \tg^{00} = (1-2\nl^2 V) \left( \frac{h' - K_+}{\Omega^2} \right) \left( h' - K_- \right), \, K_\pm = \frac{\pm I + 2 \nl V}{1-2\nl^2 V}  
\end{equation}
at $\scri^+$ is guaranteed by the form of $h'$ in eq. (\ref{eq:matching_condition}), since 
$\frac{1-2M/r}{1+2M/r} K_+ =:1+\frac{1}{r^2} G  \buildrel {r\to \infty} \over \longrightarrow 1 $ in asymptotically flat spacetimes \footnotemark and we obtain
\begin{equation}
   \om \tg^{00} =   (1-2\nl^2 V) \frac{1+2M/r }{1-2M/r } \left( -H L - \frac{1}{\tr^2} G  \right) \left( h' - K_- \right).
\end{equation} 
\footnotetext{
$\frac{1-2M/r}{1+2M/r} K_+ =1 + \frac{4M}{r^2} ( \nl_1 + V_1 ) + \mathcal{O}(r^{-3} ) $ and in Kerr $\frac{1-2M/r}{1+2M/r} K_+ =1 - \frac{4M}{r^3} ( a^2 (n_z)^2 ) + \mathcal{O}(r^{-4} ) $.
}

The scalar curvature $\om\Ric$ of $\om g^{\mu\nu}$ is given by 
\begin{eqnarray}
   \om \Ric &=& - \frac{6}{\tr} \left(  (1-2 \nl^2 V ) N + \dO L (3 - 2 V) - \dO L \tr \, \om \gamma\, \nl  \right),  \\
   \om \Ric &=& -\frac{6}{\tr} \left(  (1-2  \nl^2 V ) N  + \dO L (3 - 4 V) \right) \,\, \textrm{for Kerr},  \\
   N        &:=&  \frac{1}{\Omega^2} (  \Omega'' r - \Omega'  ) = ( \ddO \Omega^2 \tr L - \dO ( \Omega - 3\dO \tr) )L^2,  \nonumber
\end{eqnarray}
where $\om \gamma:=\gamma/\Omega$ depends on the Kerr-Schild metric in question. In Kerr $\gamma = 2M/s$, compare with (\ref{eq:gammaKS}).
In certain cases it is useful to distinguish the conformal factor $\Phi$ and the rescaling factor $\Omega$, then $\Ric$ changes as in \ref{appendixB}. 
Besides $h'$, the conformal factor has to be picked carefully to avoid steep gradients in $\om \tg^{\mu\nu}$
and $\om \Ric$ (which appear on the rhs of eqs. (\ref{eq:waveKerr}),(\ref{eq:waveKerr2})) as shown in Fig. (\ref{fig:scal-curv-trans}) (right).

Next we derive the expressions for $\Gamma^\mu(\om \tg)$, where we first consider the conformal transformation behaviour of $\Gamma^\mu(g)$ under eq. (\ref{eq:gomega}):
\begin{eqnarray}
 \Gamma^{\mu}(\om g ) &=& \om \Gamma^\mu(g) + \om C^\mu,   \label{eq:gamma_mu_omega}       \\ 
 \om C^\mu            &=& -2 \dO L ( \om n^\mu - 2 \, \nl \, \om V\,  l^\mu), \nonumber
\end{eqnarray}
where $\om C^\mu$ is a tensor, $\om \Gamma^\mu(g) = \Gamma^\mu(g)/\Omega^2$, $\om n^\mu = ( 0, n^i )/\Omega$ and $\om V=V/\Omega.$ 

The transformation of $\Gamma^\mu( \om g )$ to compactified hyperboloidal coordinates involves the Hessian of (\ref{eq:compact}),(\ref{eq:ttilde}). We obtain 
\begin{eqnarray}
   \Gamma^\mu( \om \tg ) = \tilde{\Gamma}^\mu( \om g ) - \frac{\pp \tx^\mu}{\pp x^\nu \pp x^\lambda} \om g^{\nu\lambda} \\
\fl   \frac{\pp \tx^\mu}{\pp x^\nu \pp x^\lambda} \om g^{\nu\lambda} = \frac{1}{\Omega^2} \{ -h'' (1- 2\nl^2 V) + 2 h' \frac{1}{r} V ( 1-\nl^2 ) - 2 h' \frac{1}{r}, \\
      n^i \left( (\Omega'' r - \Omega')(1-2\nl^2 V) + \Omega' (5-2V)  \right) - 4 l^i \, \Omega'\nl V  \} \nonumber \\
\fl   \frac{\pp \tx^\mu}{\pp x^\nu \pp x^\lambda} \om g^{\nu\lambda} = \{ - L \frac{d}{d\tr} h' (1-2 \nl^2 V) + 2 h' \frac{1}{\tr} \om V ( 1-\nl^2 ) - {\color{red} 2 h' \frac{1}{\Omega \tr}}, \label{eq:H0} \\
      n^i( N ( 1- 2 \nl^2 V )  + \dO L ( 5 -2V) ) - 4 l^i \dO L \, \nl V \}. \nonumber
\end{eqnarray}

A closer look at the time component $\Gamma^0(\om \tg )$ reveals that the third term in eq. (\ref{eq:H0}) is divergent. It cancels with 
$-2 \dO L \om \tilde{n}^0 = -2 \dO L (-h'/\Omega)$ that appears through the coordinate transformation
$\om C^\mu \rightarrow \om \tilde{C}^\mu$, see eq. (\ref{eq:gamma_mu_omega}).  
We remove these terms from $\om H^\mu:=\frac{\pp \tx^\mu}{\pp x^\nu \pp x^\lambda} \om g^{\nu\lambda}$ and $\om \tilde{C}^\mu$ and denote them as $\om H^\mu_{\textrm{\tiny R}}$ 
and $\om \tilde{C}^\mu_{\textrm{\tiny R}}$, where $\om D^\mu$ is the remainder of the cancellation.
\begin{eqnarray} \label{eq:gamma_om_tg}
   \Gamma^\mu(\om \tg) = \om \tilde{\Gamma}^\mu(g)  + \om \tilde{C}^\mu_{\textrm{\tiny R}} - \om H^\mu_{\textrm{\tiny R}} + \om D^\mu, \\
   \om D^\mu := \{ 2h'L \frac{1}{\tr}, 0,0,0  \}.
\end{eqnarray}

\section{Numerical applications} \label{sec:num}

We are now ready for a numerical application of the methods described in the last section. We solve the system
(\ref{eq:waveKerr2}) using a (3+1) finite differencing (FD) code on attached hyperboloidal slices using $g$ as
in eq. (\ref{eq:matching_condition2}) and $h'$ as in eq. (\ref{eq:hext}) as most commonly used in the literature and
compare the evolution of the numerical errors.  Then we solve the system (\ref{eq:waveKerr}) using a pseudo-spectral
evolution scheme in (3+1), which allows higher accuracy than the FD code, to obtain new insights into the splitting
of the late time tails of the solution \cite{Zenginoglu:2009hd,Racz:2011qu} and confirm known results about polynomial decay rates at
finite radii and $\scri^+$. We also inspect the late-time part of the solution at the horizon, where oscillations
have been predicted \cite{Barack:1999ma,Barack:1999ya} but until now not regarded in numerical studies.

With both implementations we evolve non-stationary compactly supported initial data (ID) of the form $\psi(0,x^i)=0,\, \psi_j(0,x^i)=0$
and 
\begin{eqnarray}
   \psi_0(0,\tx^i) &=& Y^{l_0m}(\tx^i) e^{-(\tr-\tr_0)^2/(2\sigma^2)} \,\, \textrm{and of the form}  \label{eq:radial_ID} \\
   \psi_0(0,\tx^i) &=& e^{-(\tx^i-\tx^i_0)(\tx^j-\tx^j_0)\delta_{ij}/(2\sigma^2)},                   \label{eq:offcentered_ID}
\end{eqnarray}
where $\sigma,\tr_0,\tx^i_0$ are parameters and $Y^{l_0m}(\tx^i)$ are the spherical harmonics wrt $\tx^i$. We set
the Kerr parameter $M=1$ and define the local power index (LPI) to be $\textrm{LPI}([\psi]^{lm}) := \textrm{d}\,
\textrm{Log} |\, [\psi]^{lm} \,| / \textrm{d}\, \textrm{Log}\, t$, where $[\psi]^{lm}$ is the $lm$-spherical harmonic
component of the field wrt Kerr-Schild coordinates.

\subsection{Finite differencing code}

Our finite differencing code \texttt{LlamaWaveHyperboloidal} \cite{code:2013} is embedded in the \texttt{Cactus} 
parallelization framework \cite{Goodale02a}. We made use of
the extension \texttt{Llama} \cite{Pollney:2009yz} of the \texttt{Carpet} driver \cite{Schnetter:2003rb}, which handles the
domain decomposition of grids over processors and provides the required interpolation operations for boundary communication. The
\texttt{Llama} code, comparable to the multiblock code of \cite{Schnetter:2006pg}, allows to use a spherical grid with the
``inflated cube'' coordinates \cite{Thornburg:2004dv} in the angular directions, which consists of six overlapping coordinate patches
each with two angular coordinates of the general form $\rho = \arctan ( x^i/x^k ),\, i\neq k $. The finite differences are computed
of $\om\psi_0$ and $\om \tilde{F}^i$ in the radial and angular coordinates, then transformed to a common Cartesian basis and input to
the rhs of eq. (\ref{eq:waveKerr2}). The discretization of the fluxes $\om \tilde{F}^i$ instead of the evolution variables $\psi_i$
is beneficial for a long-term stable evolution.  If used on a single Cartesian grid, see for example \cite{Calabrese:2003vy} and
references therein, the form of eq. (\ref{eq:waveKerr2}) is a necessary condition for the FD scheme to conserve a certain discrete
energy norm \cite{gustafsson1995time}.  Since we compute non-Cartesian FDs and use multiple overlapping coordinate patches, we
require additional Kreiss-Oliger type artificial dissipation \cite{kreiss1973methods} to guarantees long-term stability. We use
the method-of-lines with 4th-order Runge-Kutta time integration, 6th-order FD operators and add 5th-order artificial dissipation
\footnotemark (except at the radial boundaries) to the rhs of the evolution eq. for $\psi_0$.  \footnotetext{This is different from
the 6-patch-code of \cite{Schnetter:2006pg}, where summation-by-parts FD and dissipation operations are used at grid and patch boundaries.}

\subsubsection{FD code: Comparison of matching methods }

As a non-trivial test case we evolved off-centered Gaussian ID with $\tx^i_0=(2.4,1.2,1.1),\, \sigma=1$ on the dented
$\tt$-slicing, see eq. (\ref{eq:hext}), corresponding to the solid light gray curve in Fig. (\ref{fig:coord_speeds})
(left) and on the smoother $\tt$-slicing, see eq. (\ref{eq:matching_condition2}), corresponding to gray curve
in Fig. (\ref{fig:coord_speeds}) (left). We set the radial grid spacing to $\Delta \tr=0.05$, the time step to
$\Delta \tt = 0.04$ and the number of angular grid points per patch to $N_\sigma \times N_\rho = 41 \times 41$.

\begin{figure}[t] 
  \begin{minipage}[b]{0.5\linewidth}
  \centering
  \includegraphics[width=1.0\columnwidth,clip]{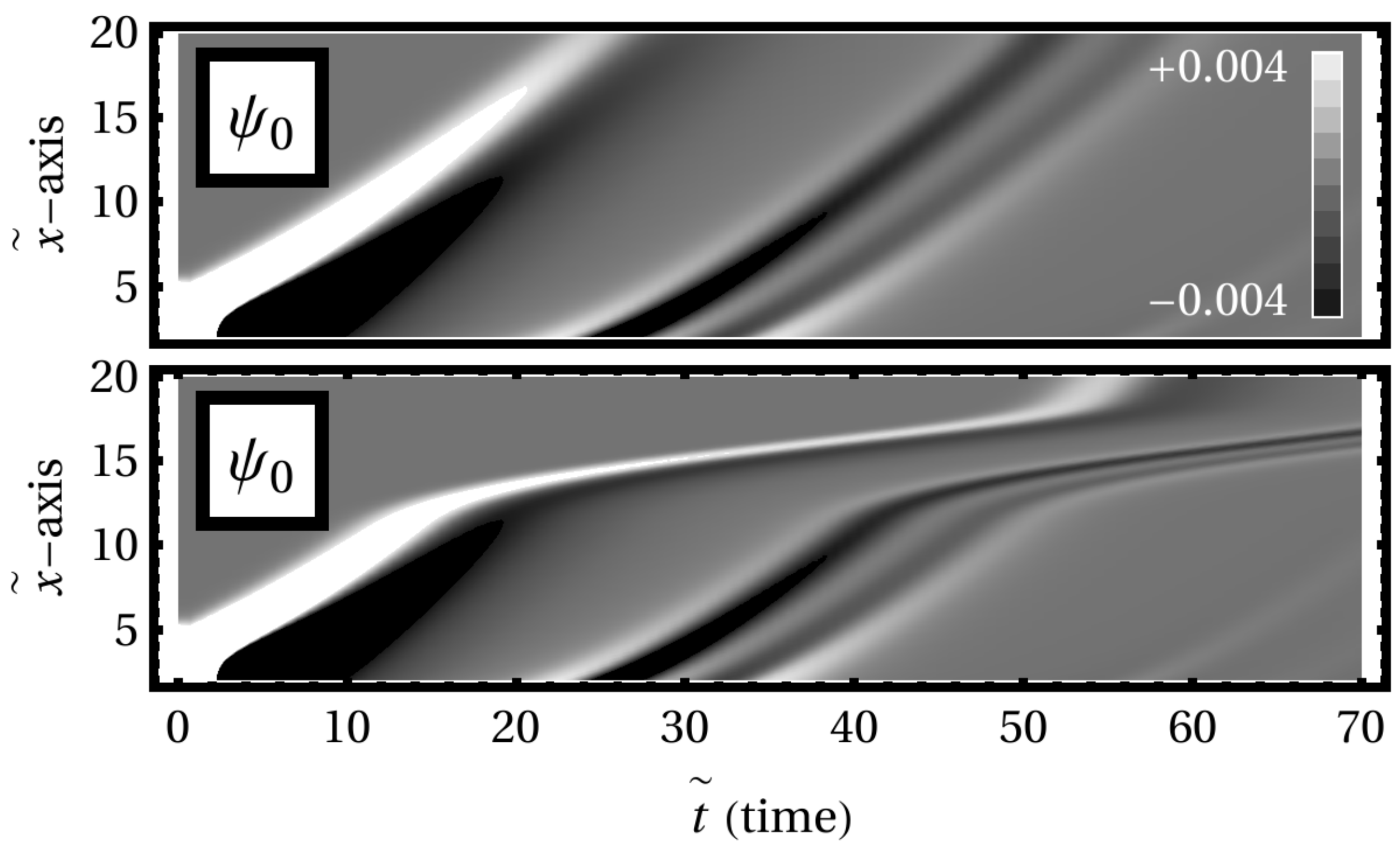}
  \end{minipage}
  \begin{minipage}[b]{0.5\linewidth}
  \centering
  \includegraphics[width=1.0\columnwidth,clip]{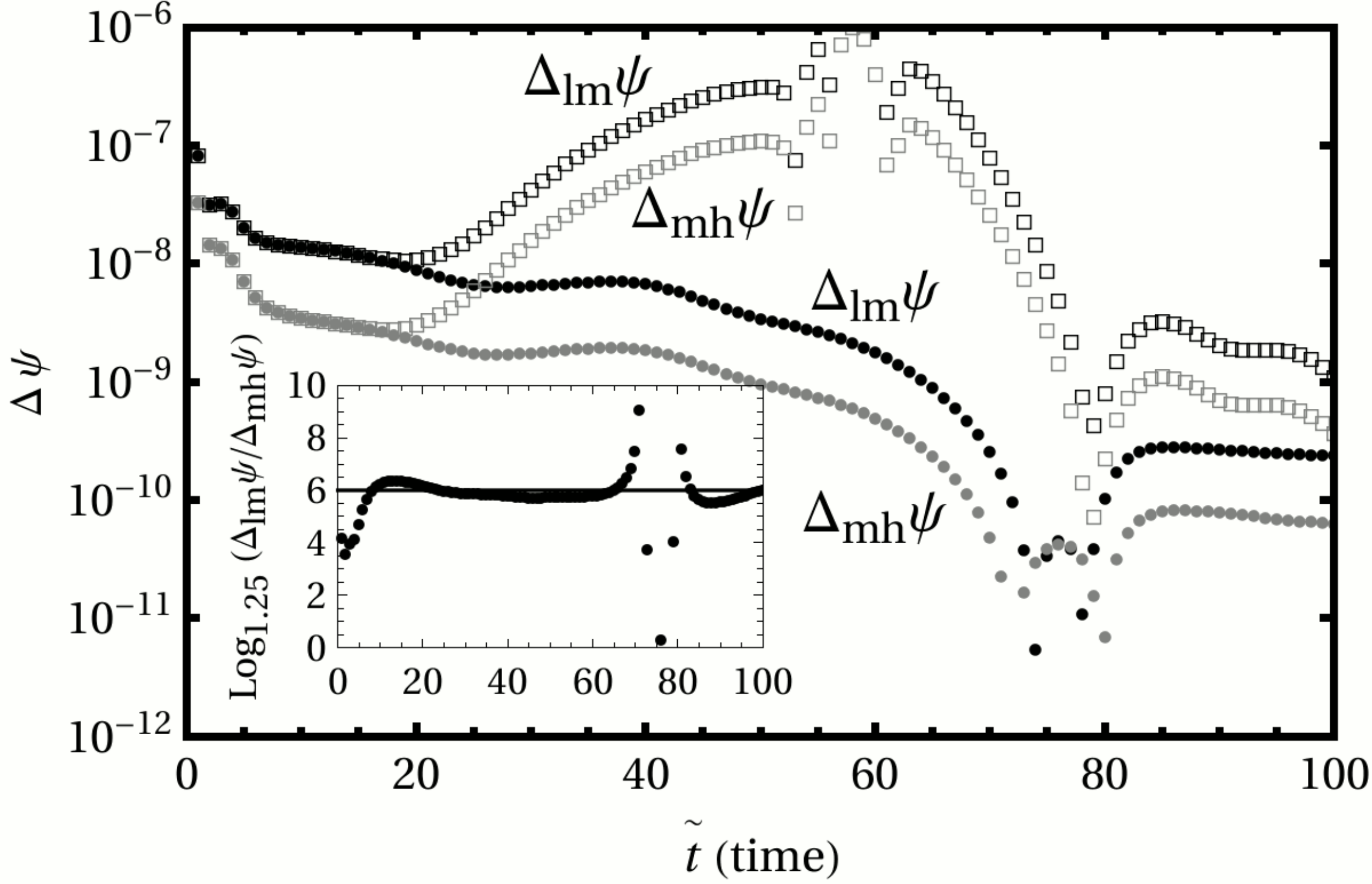}
  \end{minipage}
  \caption{Left: Evolution of $\psi_0$ along $\tx$-axis for off-centered ID on dented hyperboloidal
                 slicing (bottom) and smooth slicing (top) for Kerr $a=0.1$. Slicing parameters and
                 characteristic speeds as in Fig.(\ref{fig:coord_speeds}) (left) light gray ($f$-tanh-tan) and gray curve (eq.(\ref{eq:matching_condition2})).
                 The o. characteristic speed $c^+$ drops in the hyperboloidal domain which is apparent in the solution. \\
                 Right: Evolution of the errors of $\psi$ along $\tx$-axis for $l_0=0$ ID
                 on dented slicing (squares) and smooth slicing
                 (dots) for three resolutions $\Delta_{l,m,h} \tr=0.0625,0.05,0.04$
                 with fixed $N_\rho = 41$. Inset: Convergence rate on smooth slicing.}
\label{fig:psi0-2d-plot}
\end{figure}

In Fig. (\ref{fig:psi0-2d-plot}) the outward propagation of the initial Gaussian pulse along the $\tx$-axis
is shown. The pulse passes the transition point $\rtrans=10$ on the smooth slicing without apparent
interferences (top panel). On the dented slicing (lower panel), the pulse slows down in agreement with $c^+$
of Fig. (\ref{fig:coord_speeds}) (light gray) and leaves the numerical domain at later time. The larger
gradients in the field at a given time, due to the bump in $c^+$, cause the numerical error to be bigger as
shown in Fig. (\ref{fig:psi0-2d-plot}) (right).

\begin{figure}
  \includegraphics[width=0.5\columnwidth,clip]{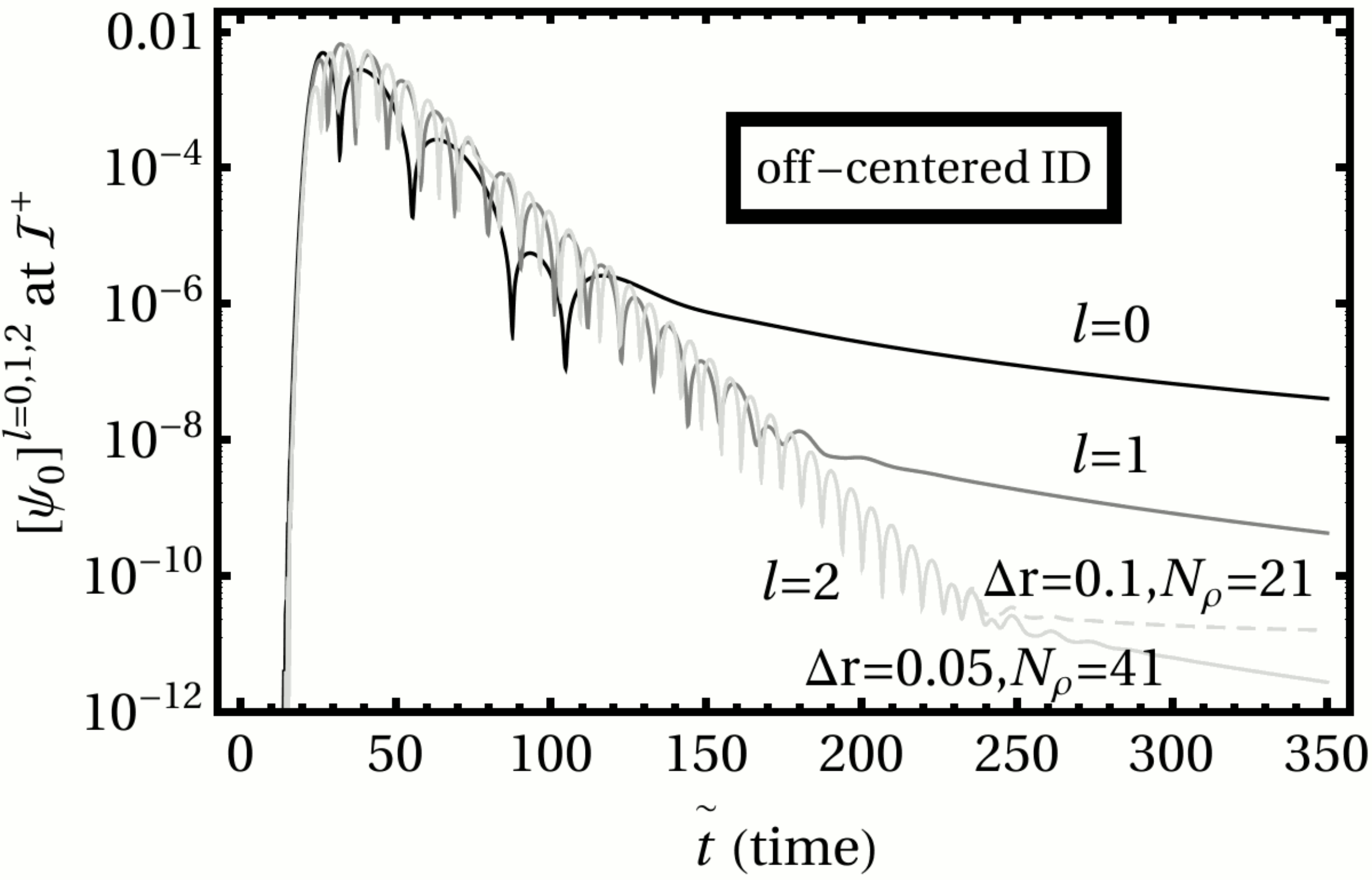}
  \includegraphics[width=0.5\columnwidth,clip]{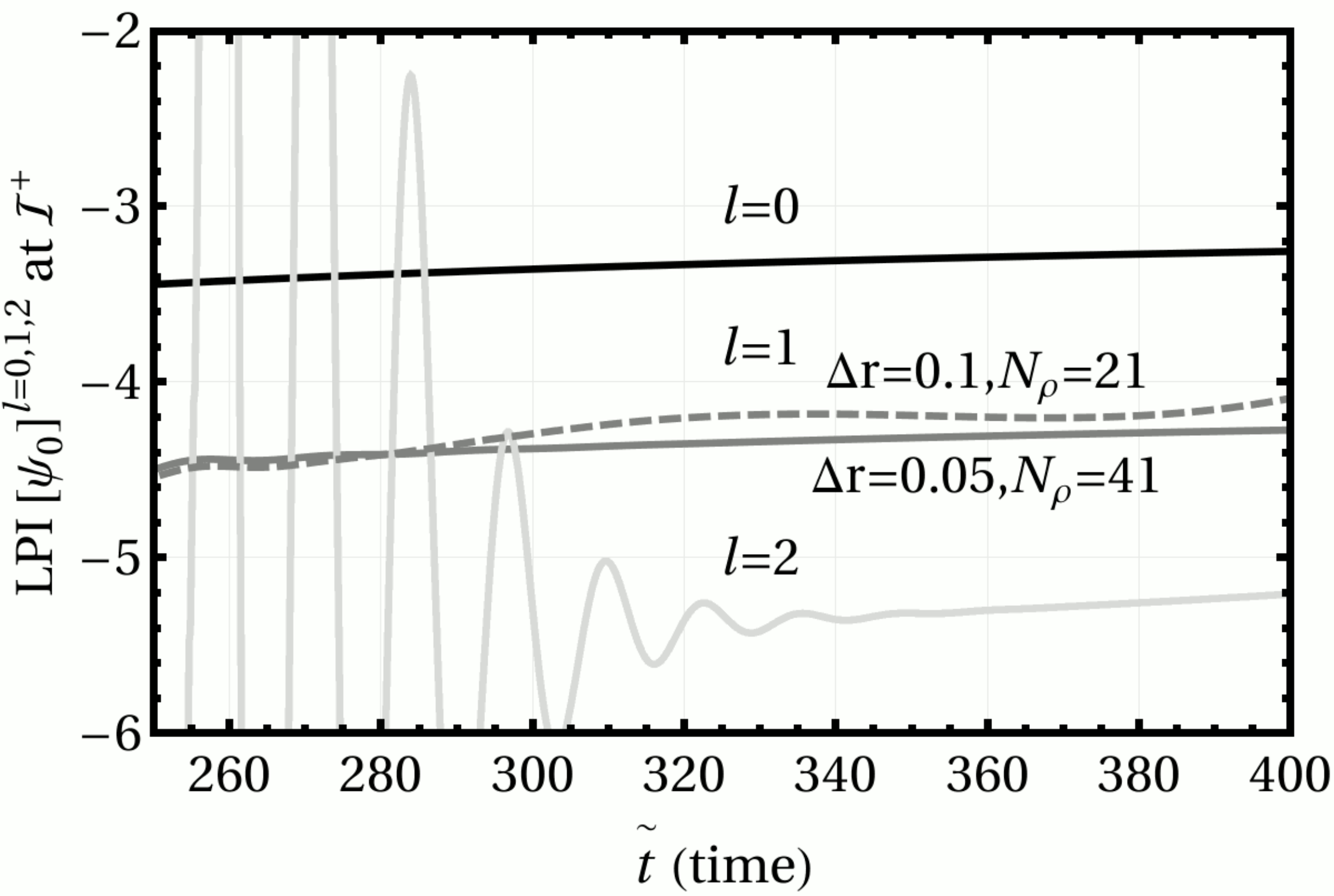}
  \caption{Evolution of the components $[\Psi_0]^{l=0,2,4}$ at $\scri^+$ for off-centered ID
           on smooth hyperboloidal slicing. Setup as in Fig. (\ref{fig:psi0-2d-plot}) (left top). The field components
           decay at late times with a power law $\sim\tt^{-n}$ as apparent in the right plot. Sufficiently high resolution 
           is necessary to see the correct power law at low amplitudes (dashed curves).}
  \label{fig:psi0-modes-at-scri-fd}
\end{figure}

The decomposition of the field $\psi_0$ at $\scri^+$ into spherical harmonics is shown 
in Fig. (\ref{fig:psi0-modes-at-scri-fd}), an exponential oscillatory decay, followed by a polynomial
decay with the asymptotic behaviour $\sim\tt^{-n}$ in agreement with the rule $n=l+2$ for $\psi$, i.e. $n=l+3$ for $\psi_0$, derived analytically
in \cite{Hod:1999rx} and confirmed in the recent numerical studies \cite{Zenginoglu:2009hd,Racz:2011qu}.
We go into more detail about asymptotic decay rates in the next section, where we use pseudo-spectral methods 
and quad machine precision. 

\subsection{Spectral code}

To resolve the late-time part of the solution higher numerical accuracy and machine precision is required than we
are able to obtain with the FD implementation described in the last section, see Fig. (\ref{fig:psi0-modes-at-scri-fd}).
For this reason we created a code that uses spectral methods \texttt{SpectralWaveKerrHyperboloidal} \cite{code:2013} 
to compute spatial derivatives and the method of
lines with 4th order Runge-Kutta time integration.  In the spectral domain the evolution variables are expanded
into real spherical harmonics $Y^{lm}$ (angular direction) and Chebyshev polynomials $C^n$ (radial direction),
i.e. into the basis of polynomials $P^{lmn} := Y^{lm}\, C^n$, where we use Gauss-Legendre,Gauss-Lobatto collocation
points, respectively. The hyperboloidal method in combination with horizon penetrating coordinates does not require
any boundary treatment and unlike the FD implementation we use hyperboloidal slices that cover the whole
numerical grid, i.e. beginning at the first, ending at the last collocation point, where we use as for the FD code the
polynomial conformal factor (\ref{eq:omega-choice}) with $k=6$, which is exactly represented in the coefficients space.
If we wanted to use a hyperboloidal domain that covers the grid only partially, we would require two spectral domains
that are joined at the transition point $\rtrans$. Since the piecewise function $\Omega$ (\ref{eq:omega-choice}) defined on a single spectral domain contains the
Heaviside function which is badly represented in the coefficients space.

The spherical harmonics and their derivatives $Y^{lm},\pp_j Y^{lm}$ are expressed in Cartesian coordinates,
similar to SpEC, see e.g. \cite{Scheel:2003vs}. They are evaluated through another basis of harmonics $\Phi^{lm}=(n^j \mathcal{N}^{[lm]}_j)^l$, where $\mathcal{N}^{[lm]}_j$
are constant complex null vectors $\mathcal{N}_i \mathcal{N}_j \delta^{ij} = 0$ labeled by $l$ and $m$, spanning the $2l+1$ harmonics
in each $l$-eigenspace. We chose the same $\mathcal{N}^{[lm]}_j$ as in \cite{Jasiulek:2009zf} such that the $Y^{lm}$ and the $\Phi^{lm}$ 
are related by a Fourier transform in each $l$-eigenspace
\begin{equation} \label{eq:ylmphilm}
\fl \quad\quad\quad Y^{lm} = B^{l m} \sum^l_{m' = -l} \Phi^{l m'} e^{-i\, m' m \,  a_l}, \quad B^{l m} = (-1)^{m} \frac{1}{l!} \sqrt{\frac{(l+m)!(l-m)!}{4\pi (2l+1)}},
\end{equation}
where $a_l=2\pi/(2l+1)$. The derivatives $\pp_j \Phi^{lm}$ take the simple form
\begin{equation} \label{eq:dphilm}
   \pp_j \Phi^{lm} = \pp_j n^i (n^k \mathcal{N}^{[lm]}_k)^{l-1}\, l\, \mathcal{N}^{[lm]}_j,
\end{equation}
where $\pp_j n^i =(\delta^{ij} - n^i n^j)/r$ and thus are tangential to the unit sphere. Given the decomposition of a function $\psi$ into the polynomial basis $P^{lmn}$ 
we obtain the tangential component of the derivative ${^{\scriptscriptstyle{||}}}(\pp_j) \psi$ through $\pp_j Y^{lm}\, C^n$
and the radial component $\pp_r \psi$ through $Y^{lm} \pp_r C^n$ (or alternatively, through a recurrence relation for the Chebyshev coefficients).
The complete partial derivative of $\psi$ is then given by
\begin{equation}
   \pp_j \psi = n_j \, \pp_r \psi + {^{\scriptscriptstyle{||}}}(\pp_j) \psi .
\end{equation}
Since we compute the rhs of eq. (\ref{eq:waveKerr}) in the physical space, we introduce an aliasing error which would cause high modes $>\lmax$ to blow up
during the evolution. For our implementation it was enough to erase the $\lmax$-components of the rhs of $\psi$ and $\psi_0$ at every step 
of the time integration to obtain a stable long term evolution and to leave the rhs of $\psi_j$ untouched.

\subsubsection{Spectral Code: Application to Late-time Decay Rates in Kerr }
A generic non-stationary compactly supported perturbation on Schwarzschild background decays at late times as
$t^{-n}\,, n=2l+3$, Price \cite{Price:1971fb}. On Kerr background the picture is more complicated. Since Kerr is not spherically
symmetric, neighbouring $l$-modes are coupled. Until recently there was a controversy
in the literature about the value of $n$ for $l\ge4$. In the simple picture the lowest $l$-mode (compatible with
the azimuthal and equatorial symmetries of the ID) generated by mode mixing during the evolution is dominating
the solution at late times. Nevertheless, analytical work by Hod \cite{Hod:1999rx}, Barack and Ori \cite{Barack:1999ma} predict a
surprising dependence of $n$ ("memory effect``) on the initially excited mode $l_0$, i.e. $n=n(l,l_0)$. By several
numerical studies \cite{Scheel:2003vs,Tiglio:2007jp,Burko:2010zj,Burko:2007ju} it was justified that both pictures
are correct and that the decay rate depends on the details of the initial value formulation, i.e. the particular
coordinates (Kerr-Schild, Boyer-Lindquist) in which the spherical harmonics are defined and the particular choice of
initial data. Therefore, we distinguish between BL harmonics $\lBL$ and KS harmonics $\lKS$ in the following through
subscripts, where necessary. Another effect that was predicted by Barack and Ori \cite{Barack:1999ma,Barack:1999ya}
are oscillations on the late-time solution at the horizon (OAH).

Recently, the use of hyperboloidal slices in Kerr allowed Zengino\u glu and Tiglio \cite{Zenginoglu:2009hd} to extend
these numerical investigation of polynomial decay rates at finite radii to future null infinity (compactly supported
non-stationary ID $l_0<5$). Very recently, Racz and Toth \cite{Racz:2011qu} presented a more detailed study for various kinds
of ID, including different fall-off properties towards null infinity up to $l_0<6$ where they investigated the
behaviour of the sub-dominant modes as well ($l$-modes with decay faster than the slowest decaying mode). They could
confirm numerically the $m$-independent rule, $n= l_0 + l +3$ for $l\ge l_0$ and $n=l_0+l+1 $ for $l< l_0$ proposed
earlier by numerical studies \cite{Burko:2010zj,Burko:2007ju} which covers the formulas derived analytically in
\cite{Hod:1999rx,Barack:1999ma}.
For the late-time tails at $\scri^+$ (TAS) the results of \cite{Zenginoglu:2009hd,Racz:2011qu}
were found to fit the rule $n=l+2$ for $l\ge l_0$ and $n=l_0$ for $l\le l_0-2$ which has been obtained analytically in \cite{Hod:1999rx}.
The authors of \cite{Racz:2011qu,Zenginoglu:2009hd} found a radial splitting of the decay rates for certain values of $l_0$, $l$ and $m$, where the decay
exponent varies for observers near the black hole, distant observers and at $\scri^+$. Moreover, the effect was found to depend on 
the harmonic index $m$ \cite{Racz:2011qu}. We observe such a splitting in our simulations as well (SPL).

As an application of the hyperboloidal method we investigated the effects OAH, TAS, SPL on the late-time tails in Kerr.
At first we check the convergence of the code for the non trivial test case of off-centered ID, see Fig. (\ref{fig:convergence}).

\begin{figure}
  \includegraphics[width=0.5\columnwidth,clip]{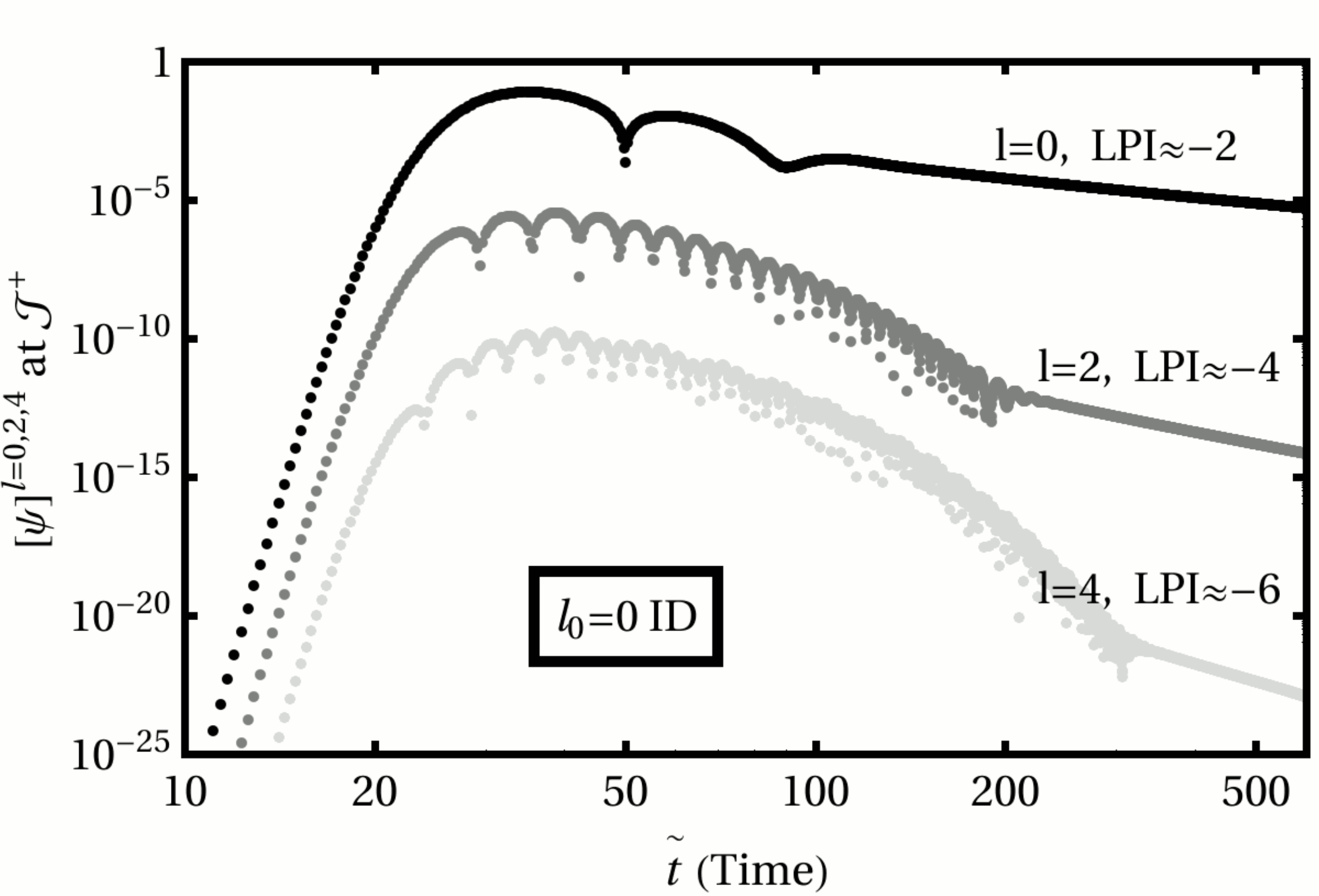}
  \includegraphics[width=0.5\columnwidth,clip]{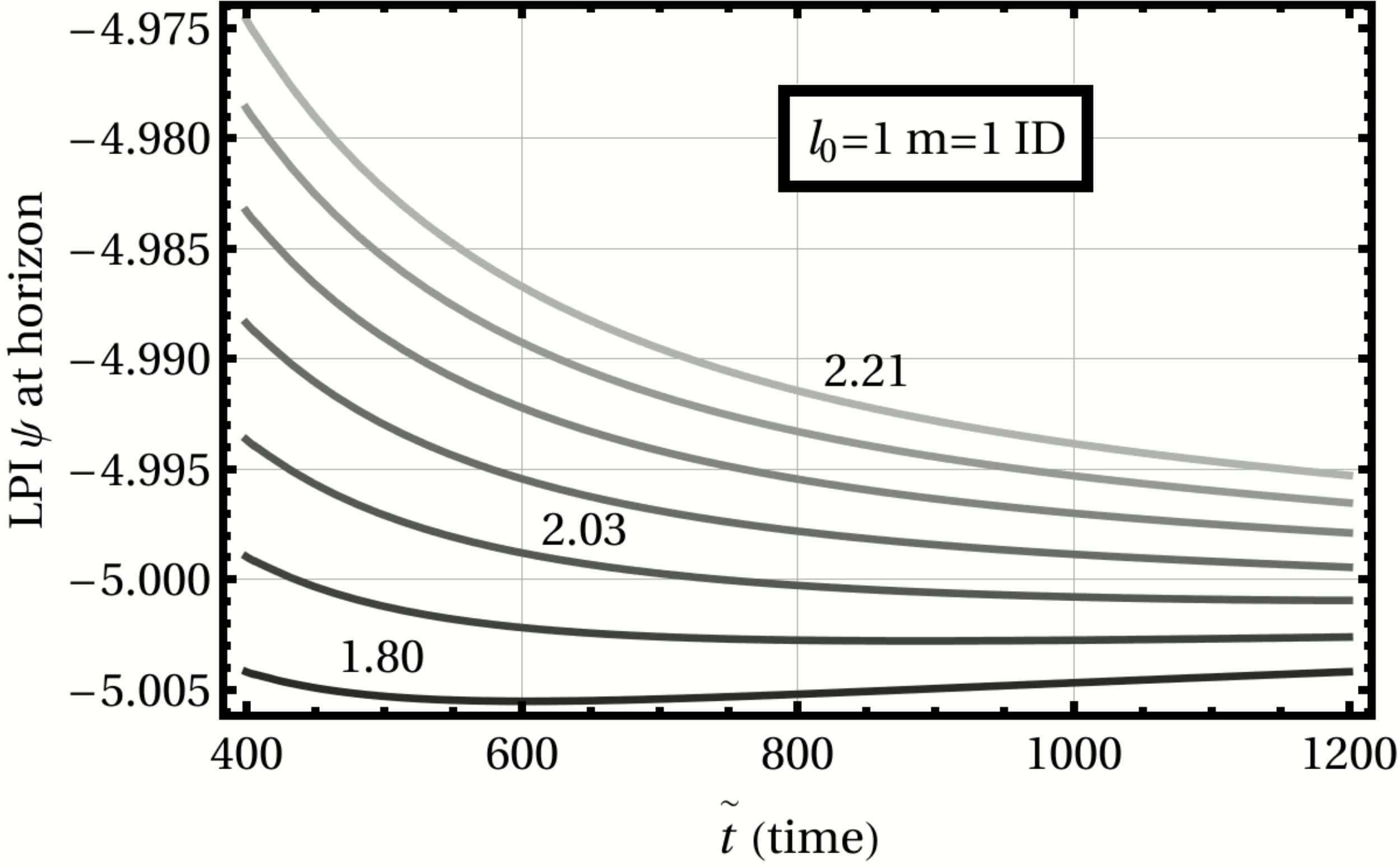}
  \caption{ Left: Field components $[\psi]^{l=0,2,4}$ at $\scri^+$ for $l_0$ ID in Kerr $a=0.5$. The decay rates are in agreement with 
                  the rule $n=l+2$. \\
            Right: LPI of $\psi$ around the horizon between $\tx^i=(1.8-2.21,0,0)$ for $(l_0=1,m=1)$ ID in Kerr $a=0.1$ with
                   $N_r=90$, $N_\theta=6$ grid points. The asymptotic decay rates approach $n=5$ in agreement with $n=l_0+l+3$.}
\label{fig:lpi_scri_horizon} 
\end{figure}

TAS: Fig. (\ref{fig:lpi_scri_horizon}) (left) confirms that the late-time tail decay rates at $\scri^+$ are in agreement with the rule $n=l+2$, 
where the evolution of the $l=0,2,4$ modes of $\psi$ for $l_0=0$ initial data is shown.

\begin{figure}
  \includegraphics[width=0.5\columnwidth,clip]{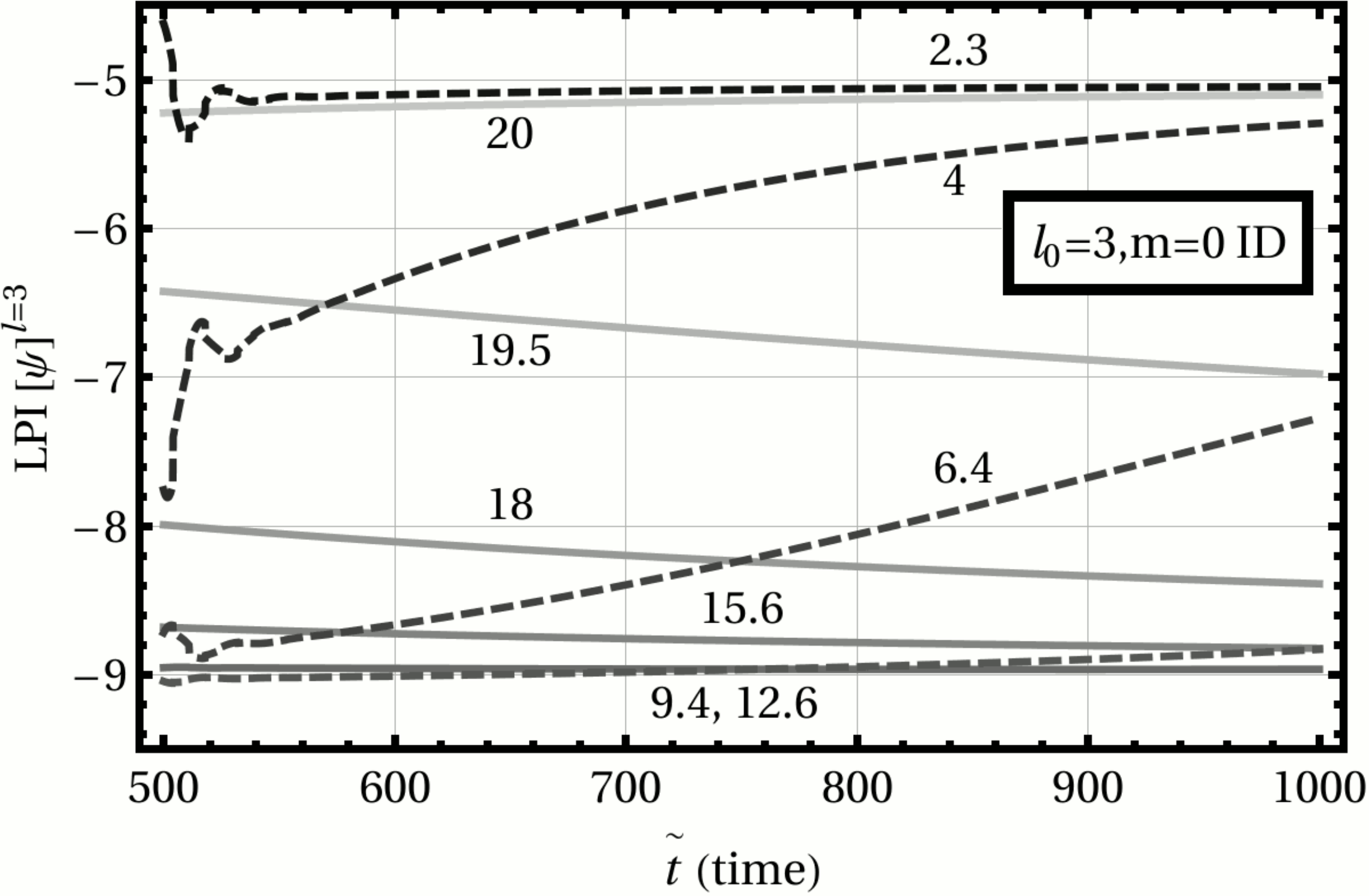}
  \includegraphics[width=0.5\columnwidth,clip]{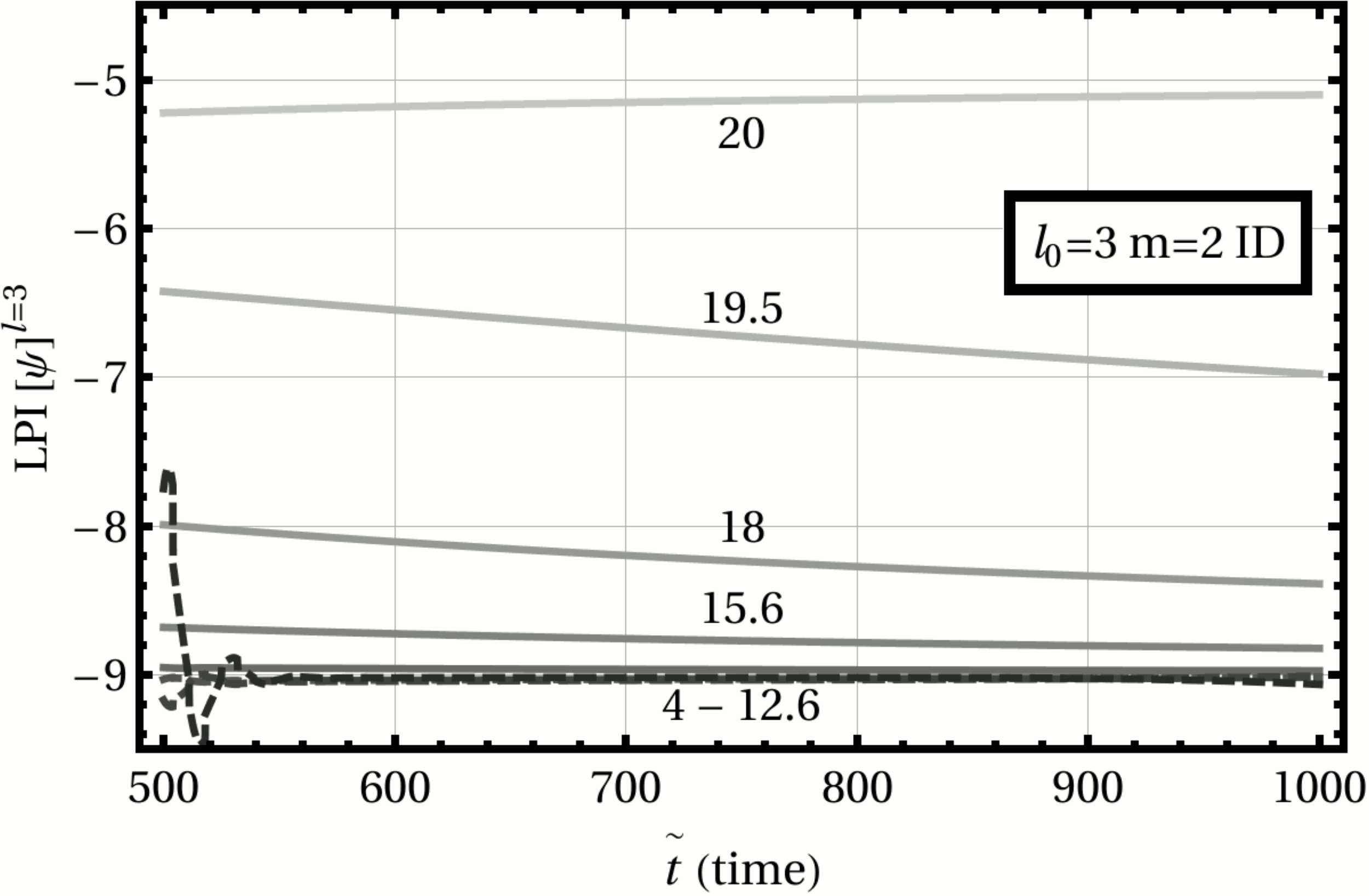}
  \caption{LPIs of $[\psi]^{l=3\, m=0}$ (left) and $[\psi]^{l=3\, m=2}$ (right) for $(l_0=3,m=0,2)$ ID at radii between $r=2-12$ (dashed lines) and $r=12-20$ (solid lines). 
           Shown is the splitting, see text, of the LPIs at different radii for $m=0$ compared to $m=2$, where no splitting appears. 
           For $N_r=90$, $N_\theta=7$ grid points and $a=0.05$.}
  \label{fig:lpi_psi3}
\end{figure}

\begin{figure}
  \includegraphics[width=0.5\columnwidth,clip]{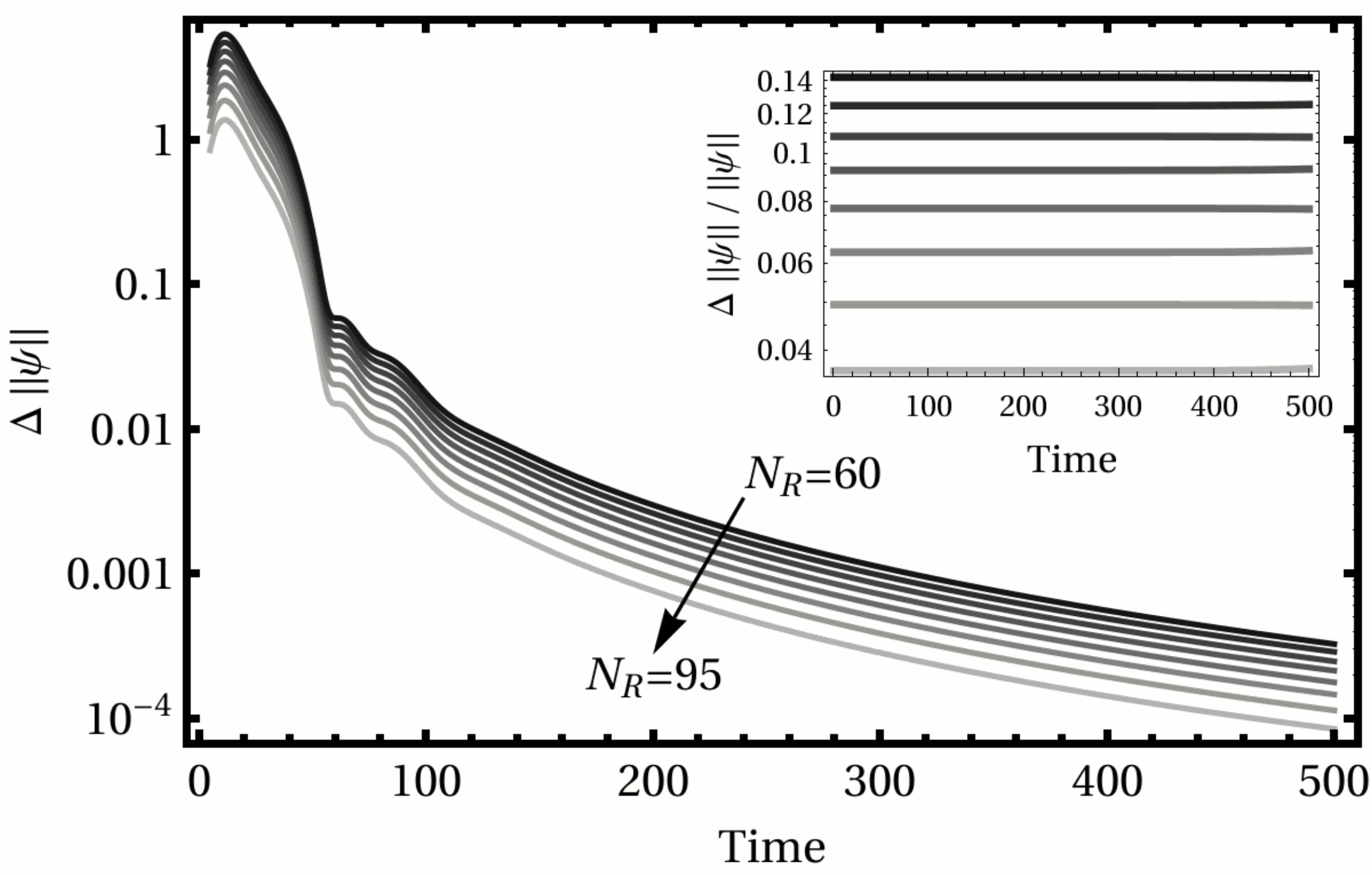}
  \includegraphics[width=0.5\columnwidth,clip]{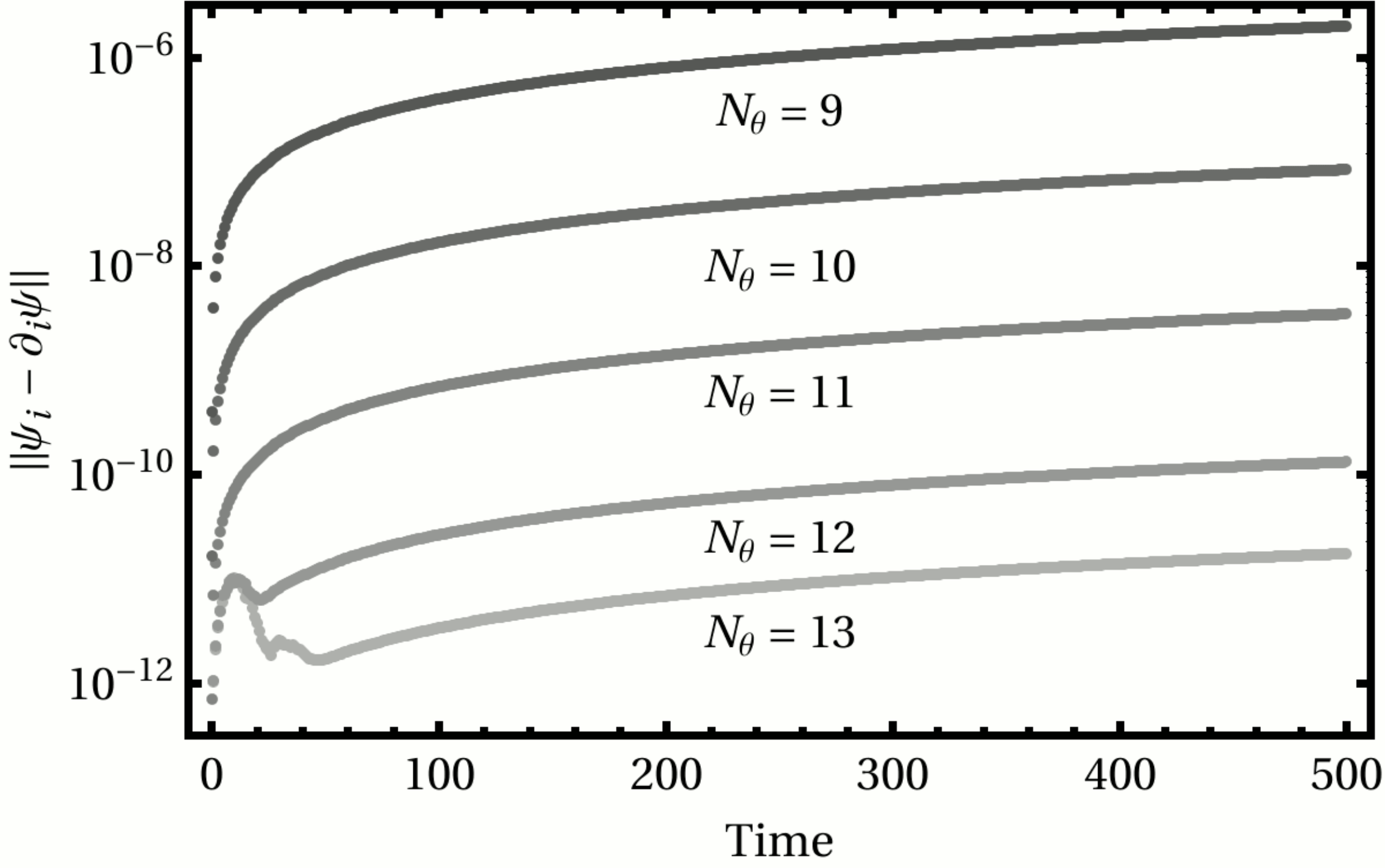}
  \caption{Left: Convergence of $L_2$-norm of $\Delta \psi$ for off-centered ID
           with $N_r=60,65,70,75,80,85,90,95$, where $\Delta \psi$ is computed wrt $N_r=110$. $a=0.9,\, N_\theta=10,\, \Delta t=0.1$ are fixed.
           Right: Convergence of constraint field 
           $\psi_i - \pp_i \psi$ at $N_\theta=13,12,11,10,9$ with fixed $a=0.9,\, N_r=65,\, \Delta t=0.1$.}
  \label{fig:convergence}
\end{figure}

OAH: In Fig. (\ref{fig:lpi_scri_horizon}) (right) we show the LPI of $\psi$ in the neighbourhood of the horizon between $\tr=1.8-2.2$ (see figure).
The field decays in agreement with the rule $n= l_0 + l +3$ but no oscillations are present. The reason is the co-rotating azimuthal
coordinate $\phi_+=\phi_{\tiny \textrm{BL}} - t_{\tiny \textrm{BL}} \Omega_+$ \footnotemark
used in \cite{Barack:1999ma,Barack:1999ya}, where the Kerr-Schild $\phi_{\tiny \textrm{KS}} = \arctan y/x$ as $\tilde{\phi}$,
which was used in \cite{Krivan:1996da}, are independent of $t_{\tiny \textrm{BL}}$. 
\footnotetext{
According to \cite{Barack:1999ma,Barack:1999ya} the field decays at late times like $\sim t^{-n}\, e^{i m \Omega_{+} v(t,x^i)} $,
where $v\sim t$ is the Eddington-Finkelstein null coordinate and $\Omega_+ = a/( 2M r_+),\, r_+ = M+(M^2-a^2)^{1/2}$.
}

SPL: As we can see in Fig. (\ref{fig:lpi_psi3}) (left) the LPI of $[\psi]^{l=3,m=0}$ for $(l_0=3,m=0)$ ID varies between $n=5$ for $2<\tr<12$ (dashed),
$n=9$ for $12<\tr<20$ (solid) and $n=5$ for $\tr=20$ at $\scri^+$. We observed the same splitting for $(l_0=3,m=1)$ ID (not shown).
This effect is not present for $(l_0=3,m=2)$ ID as shown in Fig. (\ref{fig:lpi_psi3}) (right)
and for $(l_0=3,m=3)$ ID (not shown). 
The reason is that Kerr-Schild $(l_0=3,m=0,1)$ ID as well as the examined $l$ mode contain the $\lBL=1$ harmonic which decays
with $n=1+1+3$. The ratio of the $\lBL=1$ harmonic in the $l=3$ Kerr-Schild mode decreases in the limits $a\rightarrow 0$ and
$r\rightarrow \infty$, where we observe the field decaying with $n=3+3+3$. 
It vanishes at $\scri^+$, where the BL and KS inclination coordinates agree and there the field decays like $n=3+2$.
By setting $m>1$ we can exclude $\lBL=1$ from the initial data and the splitting vanishes.

\section{Conclusion}

We applied a compactifying conformal hyperboloidal (cch) transformation to the wave operator of Kerr-Schild
metrics and derived all expressions, removed formally singular terms of the conformal wave equation for which the
numerical evaluation at the outer boundary might be problematic. The hyperboloidal slices we used in that process
let the outgoing characteristic speed invariant under the cch transformation independent of the Kerr-Schild metric
in question. This was possible by analysing the characteristic speeds of the 1st-order reduced wave equation, whereby
we obtained access to the arbitrary part $H$ of the height function derivative for which the constraints on $h'$ are
automatically satisfied (suitable asymptotic behaviour, spacelike condition). This part allows to directly set 
the outgoing characteristic speed to the desired function of radius in the hyperboloidal domain eq. (\ref{eq:matching_condition2}) and thereby to get more control on the efficiency and accuracy
of the hyperboloidal method in numerical calculations, which we demonstrated through comparison with existing approaches in
the literature. As an application we have numerically verified known decay rates of scalar test fields and their
sub-dominant modes in Kerr at the horizon, finite radii and at $\scri^+$ and obtained new insights into the $m$-dependence of the
radial splitting of the late-time decay rates of certain harmonic modes.

The asymptotics of hyperboloidal slices simplify the boundary treatment in numerical simulations and provide a
clean solution to the wave extraction problem. Moreover, the cch coordinate transformation beyond the boundary
can act like an adaptive numerical grid on the background, i.e. $H$ and $\Omega$ can be used to uniformly
spread grid points or act as a magnifier to increase the spatial or temporal resolution, say in the neighbourhood
of a localised source. In addition to providing that, the expressions presented here could lead to applications
of the hyperboloidal approach to Kerr-Schild metrics beyond Kerr.
The next step in improving the PS code would be to implement two spectral domains to be able to 
use piecewise defined conformal factors. It would be interesting to test the stability of the FD and PS code with
non-linear source terms as in \cite{Zenginoglu:2010zm} or to use a localised moving effective source \cite{Vega:2011wf}
to compute the scalar self-force on a particle orbiting a Kerr black hole and finally to extend the code to 
handle gravitational perturbations on Kerr-Schild backgrounds. The FD and PS code are available online \cite{code:2013}. 

\section*{Acknowledgments}
                                                                      
It is a pleasure to thank Badri Krishnan, An\i l Zengino\u glu, Frank Ohme, Barry Wardell, Alex B. Nielsen, Jos\'e
Luis Jaramillo, Gaurav Khanna, Leor Barack for useful discussions. I am grateful for advice from Ian Hinder and Daniela Alic on the long-term
stability of FD codes, and from Nico Budewitz for technical support on the computer cluster at the AEI. I would also like to thank 
the many developers of the \texttt{Cactus}, \texttt{Carpet} and the \texttt{Llama} code and the very useful \texttt{SimFactor}. This
work was supported by the IMPRS for Gravitational Wave Astronomy in the MPS.
                                                                     
\appendix
\section{Metric components $\om \tg^{ij}$, $\om \tg^{0j}$ and $det \om \tg_{\mu\nu}$ } \label{appendix}

The cch transformation (\ref{eq:compact}),(\ref{eq:gomega}),(\ref{eq:ttilde}) change
the metric components $g^{0j}$, $g^{ij}$ of eq. (\ref{eq:ks-metric}) to 
\begin{eqnarray}
   \om \tg^{0j}   &= \om \tilde{\eta}^{0j}   - 2 \om V\, \tl^0 \om \tl^j, &\om \tg^{ij}    = \om \tilde{\delta}^{ij} - 2 V  \om \tl^i \om \tl^j,     \\
   \om \eta^{0j}  &= -h' L n^j,                                           &\om \delta^{ij} = \delta^{ij} + \tr \dO L ( 2 + \tr \dO L )\, n^i n^j, \\ 
   \tl^0          &= l^0 - h' \nl,                                        &\om \tl^i       = l^i + n^i \tr \dO L\, \nl,  
\end{eqnarray}
where $\om V = V/\Omega$, in Kerr $\om V= \om \rbl/ \om s $ with $\om s(\tr,\tilde{z}) = \Omega^2 s(\tr,\tilde{z}) = \sqrt{ (\tr^2 - (\Omega a)^2)^2 + (2 a \Omega \tilde{z})^2 }$ 
and $\om \rbl = \frac{1}{\sqrt{2}} \sqrt{ \tr^2 - \Omega^2 a^2 + \om s  }  $. 
The determinant is 
\begin{equation}
   det\, g_{\mu\nu} = 1 \quad \rightarrow \quad det \om \tg_{\mu\nu} = 1/L^2.
\end{equation}

\section{$^\Phi \Ric$ for $\Phi \neq \Omega$ } \label{appendixB}

In certain cases it can be useful to distinguish the conformal factor $\Phi=\Omega \lambda$ and the radial rescaling factor $\Omega$,
i.e. $\lambda \neq 1$, say if the metric should be left unchanged but the numerical grid has to be adapted in some region. 
In that case $\om g^{\mu\nu} \rightarrow \omlam g^{\mu\nu}$ in eq. (\ref{eq:waveKerr2}) and $^\Phi \Ric$ becomes 
\begin{eqnarray}
   ^\Phi \Ric &=& - \frac{6}{\lambda^3} \left[ \frac{1}{\tr} \right( ( 1-2\nl^2 V ) \hat{N} + L \dot{\Phi} (3-2V) \left) - L \dot{\Phi} \nl \om \gamma   \right], \\
   ^\Phi \Ric &=& - \frac{6}{\lambda^3 \tr} \left[ ( 1-2\nl^2 V ) \hat{N} + L \dot{\Phi} (3-4V) \right] \quad \textrm{in Kerr},
 \end{eqnarray}
where $\hat{N}:= \frac{1}{\Omega^2} ( \Phi''r-\Phi' )= L^2(L_1 \dot{\Phi} + \Omega \ddot{\Phi} )\tr - L \dot{\Phi}$, $L_1:=2 \dO + \ddO \Omega \tr L $. 
The conformal transformation $\Gamma^\mu(\om \tg) \rightarrow \Gamma^\mu(\omlam \tg)$  of eq. (\ref{eq:gamma_om_tg}) is given by 
\begin{eqnarray}
   \Gamma^\mu(\omlam \tg) = \om \tilde{\Gamma}^\mu(g)/\lambda^2  + \omlam \tilde{C}^\mu_{\textrm{\tiny R}} - \om H^\mu_{\textrm{\tiny R}}/\lambda^2 + \omlam D^\mu, \\
   \omlam C^\mu  = -2 \dot{\Phi} L ( \omlam n^\mu - 2 \, \nl \, \omlam V\,  l^\mu)/\lambda^2,  \\
   \omlam D^\mu = \{ 2h' L \frac{1}{\tr} \left( \tr \dot{\lambda}/\lambda + 1  \right), 0,0,0  \}/\lambda^2.
\end{eqnarray}

\section*{References}

\bibliographystyle{unsrt}
\bibliography{main}

\end{document}